%% file: main.tex
\documentclass[lettersize,journal]{IEEEtran}
\IEEEoverridecommandlockouts
\usepackage{cite}
\usepackage{dblfloatfix}
\usepackage{amsmath,amssymb,amsfonts}
\usepackage{algorithm}
\usepackage{algorithmic}
\usepackage{graphicx}
\usepackage{hyperref}
\usepackage{textcomp}
\usepackage{xcolor}
\usepackage{xargs}                          \usepackage{subcaption}                     \usepackage{listings}
\usepackage{color}
\usepackage{dirtree}
\usepackage{multirow}
\usepackage{tabularx} \usepackage{xargs}                          \usepackage{makecell}
\newcommand{\tablestretchparam}{1.1}
\usepackage{float}
\usepackage{enumitem}
\usepackage{soul}
\usepackage{booktabs}
\usepackage{comment}
\usepackage{eso-pic}

\allowdisplaybreaks

\definecolor{dkgreen}{rgb}{0,0.6,0}
\definecolor{gray}{rgb}{0.5,0.5,0.5}
\definecolor{mauve}{rgb}{0.58,0,0.82}
\lstdefinestyle{c_code}{frame=tb,
  language=C,
  aboveskip=3mm,
  belowskip=3mm,
  showstringspaces=false,
  columns=flexible,
  basicstyle={\small\ttfamily},
  numbers=left,
  numberstyle=\tiny\color{gray},
  keywordstyle=\color{blue},
  otherkeywords={define,\#},
  commentstyle=\color{dkgreen},
  stringstyle=\color{mauve},
  breaklines=true,
  breakatwhitespace=true,
  tabsize=3
}
\lstdefinestyle{math_code}{frame=tb,
  language=C,
  aboveskip=3mm,
  belowskip=3mm,
  showstringspaces=false,
  columns=flexible,
  basicstyle={\small\ttfamily},
  numbers=left,
  numberstyle=\tiny\color{gray},
  keywordstyle=\color{blue},
  otherkeywords={define,\#},
  commentstyle=\color{dkgreen},
  stringstyle=\color{mauve},
  breaklines=true,
  breakatwhitespace=true,
  tabsize=3
}

\newcommand{\new}[1]{#1}

\begin{document}

\title{A Centralized Performance Monitoring \\ Architecture for 
Heterogeneous Multicore SoCs}

\author{
\IEEEauthorblockN{
Mohammed Sajjad Jafri, Abdur Rahman\IEEEauthorrefmark{1}, Emon Sarkar\IEEEauthorrefmark{1},\\
Mahdi Hassen\IEEEauthorrefmark{1}, Gopishankar Thayyil, Ashwin Krishna Mani, Rodolfo Pellizzoni\\[2pt]
}
\IEEEauthorblockA{
Dept. of Electrical and Computer Engineering, University of Waterloo, Waterloo, Canada\\
\{msjmoham, abdur.rahman, esarkar, mhassen, gthayyil, akmani, rpellizz\}@uwaterloo.ca
}
\thanks{\IEEEauthorrefmark{1}Equal contribution; order is alphabetical.}
}

\maketitle
\vspace*{-0.15in}
\pagestyle{headings}

\AddToShipoutPictureBG*{%
  \AtPageLowerLeft{%
    \makebox[\paperwidth][c]{%
      \raisebox{0.30in}{%
        \fbox{\parbox{\dimexpr\textwidth-2\fboxsep-2\fboxrule\relax}{\footnotesize\copyright~2026 IEEE. Personal use of this material is permitted. Permission from IEEE must be obtained for all other uses, in any current or future media, including reprinting/republishing this material for advertising or promotional purposes, creating new collective works, for resale or redistribution to servers or lists, or reuse of any copyrighted component of this work in other works.}}%
      }%
    }%
  }%
}

\input{sections/00_abstract}

\input{sections/01_introduction}

\input{sections/02_related-works}

\input{sections/03_proposed_design}
\input{sections/04_implementation}

\input{sections/05_evaluations}

\section*{Acknowledgment}
This work was supported in part by the Technology Innovation Institute, Abu Dhabi, UAE and by 
the NSERC. We would like to thank Davide Rossi and his group at the University of Bologna for providing support with the PULP platform.

\bibliography{references}

\vspace{12pt}

\end{document}

%% file: sections/00_abstract.tex
\begin{abstract}
Hardware Performance Counters (HPCs) are widely used to enable event-driven software mechanisms such as profile-guided optimization, performance analysis, and dynamic resource management in real-time systems. However, in modern embedded \new{System}-on-a-Chip (SoCs), different components - including processors, accelerators, interconnects, and memory controllers - typically implement separate and heterogeneous HPC modules. This distributed monitoring infrastructure requires multiple software interfaces and complicates the collection, synchronization, and correlation of performance data across the system.

In this work, we propose a centralized performance monitoring architecture 
to efficiently collect, correlate, and process
architectural events across multiple hardware components. Our design introduces Event Monitoring Units (EVUs) that capture and forward microarchitectural events to an Advanced Performance Monitoring Unit (APMU). The APMU integrates programmable counters and a specialized processing element to support flexible, event-driven software mechanisms. We implement our design for an AXI4-based system and integrate it into a RISC-V based SoC platform, which lacks advanced cross-component performance monitoring support. We demonstrate the effectiveness of our approach through case studies on real-time resource regulation, application profiling, and counter attribution.
\end{abstract}

%% file: sections/01_introduction.tex
\section{Introduction}
\label{sec:intro}

Modern embedded systems increasingly rely on heterogeneous computing architectures, combining diverse processors, multi-tiered memory hierarchies, and hardware accelerators - to meet growing performance demands. While this approach improves power and compute efficiency over homogeneous designs, it also significantly increases system complexity. In real-time domains, this complexity makes it particularly challenging to guarantee task execution times due to contention for shared resources such as caches, interconnects, I/O, and main memory. To support high-integrity systems on such platforms, sufficient run-time observability is essential. Commercial solutions~\cite{arm_mpam,intel_rdt,xilinx_perf_ip} provide Hardware Performance Counters (HPCs) to expose low-level architectural events. These have been widely used in general-purpose and cloud systems for profiling and bottleneck analysis~\cite{profiling_ws_comp, mdelta, maphea}, and in embedded real-time systems to enable resource regulation mechanisms~\cite{memguard,memguard_dos,memguard_kvm,mempol_journal,lat_distribution} that enforce timing isolation by monitoring resource usage and dynamically controlling access, e.g., by stalling cores\cite{memguard,BRU}.

Performance counters are nowadays available not only in processors, but also in other hardware modules such as GPU~\cite{secure_gpu, gpu_llc_cntr}, caches~\cite{ LLC_cntr, LLC_side_channel_cntr}, and memory controllers~\cite{Mem_regulation, DRAM_error_cntr}. Each set of counters observes events local to its micro-architectural domain, e.g., the number of hits and misses to a platform-level cache made by different cores, the number of open and closed row accesses to the DRAM, the request latencies on the system interconnect, etc. 
However, this decentralized approach has inherent limitations. As the counters are distributed across multiple modules, the task of reconstructing system behaviour and correlating event information in different modules becomes challenging. Each set of counters often exports a different programming interface, complicating coding efforts~\footnote{In fact, in our experience proper documentation is often missing; many hardware vendors contribute code to access HPC to the Linux \textit{perf} tool~\cite{perf_website}, whose code often represents the only source to infer HPC behavior.}. Finally, the decentralization of HPC modules also increases the software latency for accessing the counters, which can in turn affect the performance of run-time mechanisms relying on counter values~\cite{overhead_analysis,adjustable_overhead}.

In practice, run-time monitoring based on HPCs relies either on polling or on event-driven (interrupt-based) sampling. Polling is straightforward but introduces software overhead and temporal imprecision. 
\new{Interrupt-based sampling can be triggered by a periodic timer, as 
in Linux \textit{perf}'s \texttt{stat -I} mode~\cite{perf_stat_man}, or 
by a counter overflow upon reaching a configurable threshold, the 
latter requiring dedicated hardware overflow support. 
Overflow-triggered sampling reduces temporal imprecision but suffers 
from interrupt skid~\cite{tip, precise_event}, i.e., a delay between 
event occurrence and interrupt delivery that can cause event 
misattribution~\cite{sok}.} 
In either case, monitoring is typically performed on the application core itself, through CPU instructions, \new{Control and Status Registers (CSRs)}, or OS-level interfaces (e.g., Linux \textit{perf}~\cite{perf_website}), introducing overhead that may perturb the workload\new{,} a critical concern in real-time systems. Offloading monitoring to a dedicated side core~\cite{mempol_journal} reduces direct interference, but still requires software-mediated reads from decentralized PMUs, incurring inter-core communication latency and synchronization overhead~\cite{overhead_analysis, adjustable_overhead}.

To further reduce software involvement, modern processors provide hardware-supported tracing and sampling infrastructures, such as \new{Arm} CoreSight~\cite{arm_coresight} and precise sampling extensions (e.g., Intel  \new{Processor Event
Based Sampling(PEBS)}~\cite{intel_pebs} and \new{Arm} \new{Statistical Profiling Extension (SPE)}~\cite{arm_spe_overview}), which capture events in hardware buffers with lower overhead and improved temporal accuracy. 
However, these mechanisms remain largely processor-centric\new{: CoreSight's~\cite{arm_coresight} standard tracing components (e.g., Embedded Trace Macrocell (ETM), System Trace Macrocell (STM)) are designed to capture high-bandwidth instruction and data traces from application cores, and while the Advanced Microcontroller Bus Architecture (AMBA) Trace Bus (ATB) can in principle carry traces from other IPs, practical deployments overwhelmingly target core-level tracing; precise-event extensions such as Intel PEBS~\cite{intel_pebs} and Arm SPE~\cite{arm_spe_overview} are similarly tied to the processor pipeline.} As a result, they do not natively support unified, low-latency correlation of events across heterogeneous system components such as caches, interconnects, and memory controllers, leaving system-wide observability dependent on multiple decentralized interfaces.

Event-driven run-time mechanisms in real-time systems require timely, low-overhead, and system-wide observability to enforce temporal isolation and resource regulation. 
Consequently, there is a need for a monitoring infrastructure that (i) minimizes software involvement, (ii) provides standardized access to events from multiple hardware components, and (iii) enables low-latency correlation and processing of performance data in a unified manner. To these ends, in this work we propose a centralized performance monitoring architecture. We provide an infrastructure where architects can instantiate custom monitoring units, called EVent Units (EVUs), in individual system components, like processors, caches, memory controllers, etc. These units are responsible for gathering architectural events and transmitting them to an Advanced Performance Monitoring Unit (APMU). The APMU collects and processes the received information:
it comprises a set of configurable smart counters that can perform arithmetic and logical operations on the received event information, and a specialized instruction processor that 
executes
software programs on the counter data.
Our major contributions are: 

\begin{itemize}
    \item \new{We propose a centralized architecture that enables unified collection of architectural events from multiple heterogeneous System-on-a-Chip (SoC) components. Our architecture is based on a standardized EVU-APMU interface to enable interoperability among event-generating components and to simplify integration of existing performance monitoring units (Section~\ref{subsec:interface}).}

    \item \new{We design an APMU for event aggregation, filtering, and counter updates. The APMU contains a dedicated processing element (APMU-PE) for counter processing and correlation, eliminating the software overhead, interrupt-induced perturbations, and temporal 
    skew that affect conventional HPC-based monitoring mechanisms executing on application cores (Section~\ref{subsec:apmu}).} Our design provides a standardized software interface for accessing counters, ensures low-latency event collection and processing, and is both flexible and configurable.

    \item \new{To prove the feasibility of our architecture, we implement in hardware description language a 
    prototype APMU alongside multiple EVUs: (i)~an AXI4 Snooping Unit (SPU), an interconnect-level 
    EVU that monitors AXI4 transactions on the system interconnection~\cite{axi}, and (ii)~a CVA6 Event Unit (CVA6-EVU) 
    that extends the native performance monitoring of the RISC-V CVA6 
    core by exposing microarchitectural events and Program Counter (PC) 
    markers for fine-grained event attribution and application progress 
    tracking (Section~\ref{sec:impl}).}
            We \new{further} integrate the proposed APMU and EVUs into an existing RISC-V based multicore platform lacking native advanced performance monitoring support~\cite{riscv_vol_1, riscv_pmu}, and validate their functionality, resource overhead, and runtime behavior on an FPGA-based emulation platform (Section~\ref{subsec:hw_eval}).

    \item Finally, we demonstrate the applicability of our architecture through two case studies: (i) a latency-based resource regulation mechanism, and (ii) fine-grained application profiling and counter attribution using PC markers (Sections~\ref{sec:regulation} and~\ref{sec:profiling}).
\end{itemize}

%% file: sections/02_related-works.tex
\section{Background and Related Works}
\label{section2:related_works}

\subsection{Hardware Performance Monitoring Counters (HPC)}
\label{subsec:HPC}

Many of the popular \new{Commercial Off-The-Shelf (COTS)} and open-source computer architectures and hardware frameworks support HPCs. On the industrial front, both Intel and Arm support HPC counters in their processors \cite{arm_a72, intel_pmu}. HPCs also form the base for many performance monitoring frameworks, such as Arm's Memory System Resource Partitioning and Monitoring (MPAM) \cite{arm_mpam} and Intel's Resource Director Technology (RDT) \cite{intel_rdt}. In the open-source domain, RISC-V~\cite{riscv_vol_1} has adopted the “Zicntr” and “Zihpm” extensions \cite{riscv_pmu}, that propose a set of performance counters: CYCLE, TIME and INSTRET, which count the number of clock cycles elapsed, wall-clock real time, and the instructions retired, respectively. Apart from these, the extensions also propose 29 additional 64-bit HPCs that can be designed to count platform-specific hardware events. Unfortunately, the current list of performance monitoring events in the CVA6 family~\cite{riscv_cbqri, riscv_ssqosid}, which is the RISC-V processor family used in our evaluation platform, is lacking. Events such as number of writebacks by the data cache and unconditional jumps, are missing. \new{Furthermore, 
the RISC-V HPC specification does not mandate support for counter overflow interrupts, limiting the implementation of event-triggered sampling mechanisms on many RISC-V implementations.}
Another major limitation of RISC-V performance counters is that they are only accessible by the processor itself. This means that a core cannot read the counters of another core directly, which makes the execution of a multicore event-based mechanism difficult.

\noindent\textbf{HPC Software Support.} Core-level HPCs often vary significantly across processor generations, with different microarchitectural events supported on specific models. In some architectures, counters are not accessible at lower privilege levels, requiring traps to higher privilege modes. Combined with the limited number of available counters, this makes direct access by application programmers uncommon on general-purpose platforms. Instead, on Linux, the \textit{perf\_event}~\cite{linux_perf_events} kernel-level abstraction layer is used to interact with HPCs, and many libraries and tools~\cite{papi, Vtune, perfutility} rely on it. Through \textit{perf\_event}, applications can access performance data via sampling or explicit polling mechanisms, each involving trade-offs between overhead and temporal precision. \new{Concretely, the \textit{perf} tool built on \textit{perf\_event}
reports counters either as totals over a workload (\texttt{perf stat})
or as periodic snapshots at a configurable interval, with a minimum
interval of $1\,ms$ in current upstream kernels
(\texttt{perf stat -I}). At each interval the kernel reads per-CPU counter state and user-space retrieves the deltas on cores that may also be running application threads. System-wide measurement (\texttt{-a}) is implemented as one file descriptor per monitored CPU, with software-side aggregation of the per-core results, introducing additional read overhead and temporal skew across the per-core reads~\cite{perf_stat_man}.  PAPI~\cite{papi} provides a portable cross-platform PMU API but inherits these costs, and on platforms where the underlying hardware lacks counter-overflow interrupts
it falls back to a software-emulated
handler driven by a periodic kernel timer. In all cases, these mechanisms execute on the cores
being monitored, imposing interference that is at odds with the timing-predictability requirements of real-time workloads.} \new{Prior work \cite{precise_event} claims that even with overflow interrupt based specialized statistical profiling tools such as Intel PEBS, one may still suffer from extreme imprecision.} While mature software stacks exist for core-level PMUs, similar abstractions are largely absent for performance counters located in other system components. \new{We note that modern x86 server platforms do expose \emph{uncore} PMU counters that monitor shared resources such as last-level cache slices, integrated memory controllers, and on-die interconnects via Linux \texttt{perf}, PAPI~\cite{papi}, or vendor performance tools such as the HPE Cray uncore subsystem~\cite{cray_uncore}. However, these are organized per socket and can only be programmed from a single designated CPU per socket; event encodings differ from core HPCs and vary across socket generations, meaning cross-component correlation still requires the application to manage multiple distinct event families from specific cores. On embedded platforms, which is our target domain, no analogous uncore PMU exists, and counters in components such as platform caches, accelerators, and interconnects are typically not visible through \texttt{perf} at all.}
For instance, GPU performance counters are typically accessed through vendor-specific profiling frameworks (e.g., NVIDIA Nsight~\cite{nsight_systems}), while counters in memory controllers, interconnects, and platform caches are often exposed through proprietary drivers, memory-mapped registers, or undocumented interfaces. Consequently, there is no standardized interface comparable to \textit{perf\_event}~\cite{linux_perf_events} kernel subsystem or PAPI~\cite{papi} that provides a unified access model for heterogeneous counters across CPUs, caches, interconnects, and memory subsystems. \new{Similar to \emph{uncore} support, our architecture monitors architectural events from heterogeneous components; however, we centralized event collection on a dedicated processing element to export a unified counter access interface and improve event correlation.}

\new{\noindent\textbf{HPC Usage: Off-line Approaches.}}
HPCs provide fine-grained visibility into microarchitectural events such as instruction execution, cache behavior, branch prediction, and memory accesses~\cite{hiddenpmu, tintin}. Prior work using HPCs can be broadly categorized into off-line and on-line techniques based on how counter data is consumed. Off-line approaches collect counter data during program execution and analyze it post-mortem to improve system understanding and performance, including performance debugging and software diagnosis~\cite{debug2, debug3}, workload profiling and program optimization~\cite{opt1, opt3, mdelta, maphea}, power and energy analysis~\cite{ power3}, and hardware characterization and timing analysis~\cite{Response_Time_Analysis, multicore_contention}.

\new{Attributing performance-counter measurements to individual functions or code regions conventionally requires intercepting function entry and exit in software. The most direct approach is compile-time instrumentation: with GCC's \texttt{-finstrument-functions}~\cite{gcc_instrument}, the compiler inserts calls to user-supplied hooks at function entry/exit, and the hooks then snapshot the counters of interest, e.g., through \texttt{PAPI\_read}~\cite{papi} or direct CSR reads. Linux \emph{uprobes}~\cite{uprobes_doc, uprobes_lwn} provide dynamic attribution on unmodified binaries: the kernel replaces the probed instruction with a breakpoint, so that each execution of the probed address traps into the kernel and runs the attribution handler. 
Although the on-disk binary is untouched, the executed instruction stream is modified. In both, the cost is paid on the application core and scales with the product of invocation frequency and handler length.}

\new{\noindent\textbf{HPC Usage: On-line Approaches.}}
\new{In contrast to off-line approaches,}
on-line techniques
rely on real-time observation of HPC measurements to drive dynamic decisions during execution, making them inherently 
\new{sensitive to counter-access latency and temporal skew.}
Representative examples include run-time resource management and adaptive scheduling~\cite{caladan, dna, dcat} and enforcement of usage thresholds for shared hardware resources~\cite{muticore_WCET}. Recent works further demonstrate that HPC variations can be leveraged for real-time detection of transient execution and side-channel attacks~\cite{hiddenpmu, side_channel}, and a comprehensive survey highlights the effectiveness of HPC-based monitoring combined with machine learning for resilient and real-time threat detection~\cite{survey_hardware_based_malware}. \new{This trend is mirrored in industrial deployments: Intel's Threat Detection Technology (TDT)~\cite{intel_tdt} drives runtime ML-based detection of cryptojacking and ransomware from CPU HPCs, and notably offloads the inference workload to the integrated GPU so that the analysis does not steal cycles from the protected application cores. This pattern, leveraging HPCs on-line while moving the processing off the application core, directly motivates the centralized, offloaded event-processing architecture we propose in this work.} \new{However, our APMU is designed as a general event processing unit that can be employed for diverse use cases, rather than targeting a specific application as TDT.} 

In the real-time domain, various on-line mechanisms based on HPC have been proposed to regulate accesses to shared hardware resources. 
MemGuard~\cite{memguard} implements a memory-bandwidth reservation system to prevent undue interference between cores contending for main memory access. MemGuard assigns a maximum budget in number of memory requests per regulation period to each core. If a core exceeds its budget during a period, MemGuard halts the core until the beginning of the next regulation period\new{; this is achieved 
by configuring the core's HPC to raise a hardware overflow 
interrupt once the budget threshold is reached.} 
 Because the budget is refreshed by triggering \new{a timer} interrupt, due to overhead considerations the minimum regulation period is set to 1 $ms$. 
Among works extending the original MemGuard approach, \cite{memguard_dos} notes that contention for access to a shared LLC can also significantly affect the cores, and thus applies MemGuard to reserve LLC bandwidth rather than main memory bandwidth.

A different approach based on periodic sampling is presented in~\cite{multi_core_mem_util, lat_distribution, mempol_journal}. In particular, MemPol~\cite{mempol_journal} exploits the heterogeneous architecture common in \new{Arm} SoCs to execute the regulation logic on a smaller side-core, rather than on the application cores themselves. 
\new{Arm CoreSight~\cite{arm_coresight}, in addition to its trace components (ETM, STM), also provides a debug access interface that allows external, core-independent access to processor state and performance counters. MemPol exploits this debug access interface for cross-core memory regulation~\cite{mempol_journal}, using it to poll PMU counters and 
halt cores externally without requiring on-core software. However, 
this access path incurs non-trivial per-transaction latency, and being inherently polling-based, 
cannot react immediately to threshold crossings, leading to 
overshoot~\cite{mempol_journal}. Our work targets this same goal of 
core-independent, cross-component visibility, but does so through 
a dedicated hardware event-packet interface rather than software 
polling over a debug access port, eliminating per-access 
latency and reducing threshold-detection delay.}

\subsection{Hardware Mechanisms for Resource Contention}

Finally, 
several works have proposed novel hardware solutions, implemented either at the architectural simulator or RTL level, to directly track and manage resource contention. In particular, in~\cite{max_contention, space_riscv, tracking_mem_contention,  tracking_coherence_contention}, various hardware contention units are proposed, which track the amount of interference that a core suffers from every other core in the system. This includes complex resources such as coherent interconnections~\cite{tracking_coherence_contention} and DRAM main memory~\cite{tracking_mem_contention}.
We argue that such works are orthogonal to our approach; the discussed mechanisms could be realized as custom EVUs and the related contention information transmitted to our APMU. This could allow, for example, the implementation of smart regulation policies by combining information about contention across multiple hardware resources, similarly to how we exploit events for both LLC and main memory in Section~\ref{sec:regulation}. 

Rather than relying on software-based approaches for memory regulations, some recent work have introduced custom hardware units. A Bandwidth Regulation Unit (BRU) that monitors and enforces bandwidth limits for access to LLC has been proposed in~\cite{BRU} and later extended in~\cite{per-bank} to multi-bank caches. It employs a MemGuard-like replacement policy with a period of 400 cycles. 
The authors of~\cite{memcore} leverage the specific capability of AMD Ultrascale+ heterogeneous SoCs to snoop coherence traffic between LLC and main memory through FPGA logic to implement a hardware main memory bandwidth regulator. The implementation is based on a modified token bucket~\cite{token_bucket}, with a lowest employed accounting period of 1 $\mu s$ (1,200 cycles of the \new{Arm} application cores), and the cores are halted through CoreSight as in~\cite{mempol_journal}. In Section~\ref{sec:regulation}, we show that our APMU-based regulation loop can run at comparable frequency, while retaining the flexibility of a software implementation and correlating information on access to both LLC and main memory. We further remark that the goal of our proposal is to introduce a framework for efficient event correlation and processing that can be used in diverse SoCs and across the range of HPC applications, and not just for memory regulation.

%% file: sections/03_proposed_design.tex
\section{Challenges and Design Overview} 
\label{sec:design}

Following the discussion in Sections~\ref{sec:intro} and \ref{section2:related_works}, we argue that existing HPC infrastructures in modern SoCs suffer from three fundamental limitations:

\new{\textbf{(C1)}} \textbf{Monitoring overhead and interference.}
Conventional HPC collection relies on software mechanisms such as interrupts, syscalls, polling, or application-core reads. These introduce execution overhead, additional memory traffic, and timing perturbations to the workload under observation. Even when monitoring is offloaded to a side core, software-mediated reads of distributed counters still incur latency and synchronization overhead.

\new{\textbf{(C2)}} \textbf{Correlation and temporal skew.}
Performance counters are distributed across heterogeneous hardware modules (cores, caches, interconnects, memory controllers). Reconstructing system-wide behavior requires multiple software reads at different time instants, leading to temporal skew, event misattribution, and increased latency when correlating cross-subsystem events at runtime.

\new{\textbf{(C3)}} \textbf{Heterogeneous interfaces and fragmented tooling.}
As discussed in the previous section, while unified abstractions (e.g., \textit{perf\_event}) exist for core PMUs, counters in other subsystems are typically exposed via MMIO registers, proprietary drivers, or vendor-specific toolchains. This lack of a unified access interface increases integration complexity and counter-access latency for system-wide monitoring.

\begin{figure}[ht]
    \centering
    \includegraphics[width=0.48\textwidth]{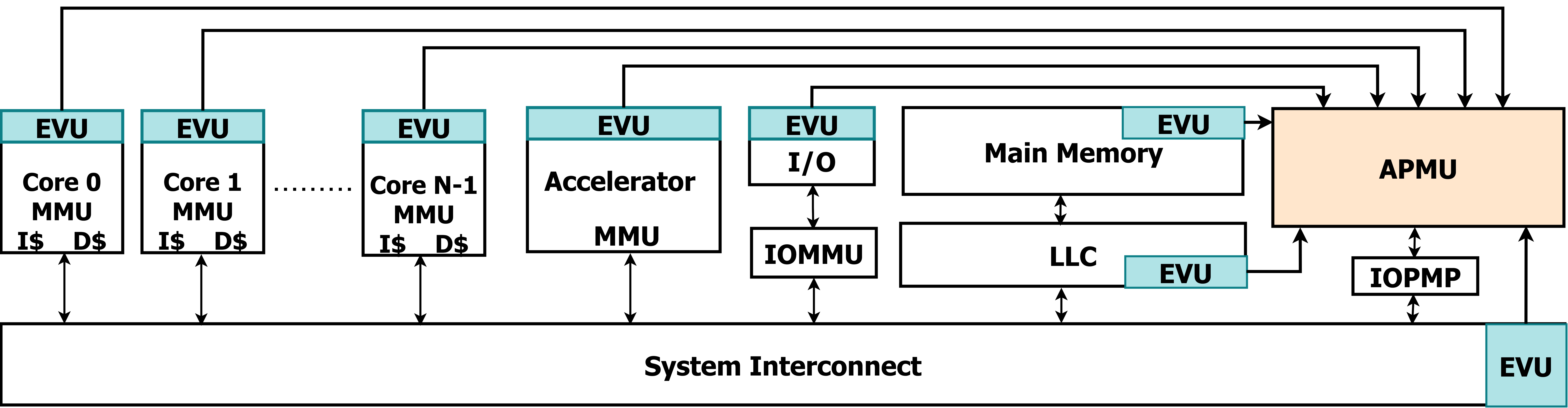}
    \caption{A generic multicore platform with our proposed centralized performance monitoring infrastructure.}
    \label{fig:mc_plat}
\end{figure}

To address these challenges, we propose a centralized performance monitoring infrastructure composed of a set of distributed EVUs and a central APMU, as illustrated in Figure~\ref{fig:mc_plat}. Each EVU is placed close to a hardware IP block (e.g., cores, interconnects, memory controllers) to capture local architectural events and transmit structured event packets to the APMU through a standardized EVU-APMU interface. This design enables low-latency event collection while maintaining interoperability across heterogeneous IPs.

A key design principle of the proposed architecture is the decoupling of event generation from event processing. The APMU integrates programmable counters and a dedicated processing element (APMU-PE) that executes monitoring and control logic directly within the monitoring subsystem, effectively acting as a specialized side core for performance analysis. By offloading counter aggregation, filtering, and decision-making to the APMU-PE, the proposed approach avoids executing monitoring routines on application cores, thereby minimizing runtime interference and preserving timing predictability in real-time workloads.

\new{A direct consequence of this principle, which shapes the software stack described in Section~\ref{sec:sw_stack}, is a \emph{load-and-run} programming model: monitoring policies are authored as self-contained components, compiled with the APMU-PE toolchain, and deployed onto the APMU-PE, either statically at boot or dynamically at run time rather than being expressed as application code or kernel modules. All event filtering, correlation, and control logic lives on the APMU-PE. This separation is what distinguishes our approach from conventional stacks such as Linux \texttt{perf}/\textit{perf\_event} or PAPI, in which monitoring code is either compiled into the application, patched into its binary, or executed by the OS kernel on the same cores that run the workload under observation.}

This centralized and side-core-based design directly addresses the limitations of existing HPC infrastructures. First, isolating event processing within the APMU eliminates interrupt and syscall-induced overheads on application cores~\new{\textbf{(C1)}}. Second, the EVU-APMU interface enables synchronized collection of events from multiple subsystems, reducing temporal skew and improving cross-component event correlation~\new{\textbf{(C2)}}. Third, the unified hardware and software interface provided by the APMU abstracts the heterogeneity of component-level counters, simplifying the development of system-wide, event-based mechanisms such as resource regulation, profiling, etc.~\new{\textbf{(C3)}.}

Section~\ref{sec:infra_design} presents the architectural design of the proposed infrastructure, while Section~\ref{sec:impl} describes its hardware implementation and integration into a RISC-V based platform.
\section{Infrastructure Design} \label{sec:infra_design}

\subsection{Event Unit (EVU) and EVU-APMU Interface} \label{subsec:interface}
Each EVU encapsulates a local view of the system, abstracting component-specific signals into a unified event reporting model, and it communicates with the APMU through a standardized interface. The EVU-APMU interface is designed as a standard interface specification, analogous in spirit to \new{Arm}’s AMBA AXI~\cite{axi} protocol. Its purpose is to enable interoperability among event-generating IP blocks and a unified APMU framework, regardless of vendor or implementation. Each EVU is connected to a dedicated port on the APMU, enabling transmission of multiple events per clock cycle depending on the interface width and protocol. The EVU-APMU interface is divided in two layers: (a) logical layer, and (b) physical layer.

The logical layer defines the semantics of the information transmitted within an event packet. Each packet comprises three fields: (i) Event ID, (ii) Event Info, and (iii) Source ID. The Event ID uniquely identifies the type of event observed by the EVU. The Event Info field carries optional metadata associated with the event instance, such as transaction size, latency, or other event-specific attributes. The Source ID identifies the originating IP, processor execution mode, or agent responsible for triggering the event. For instance, an EVU monitoring a platform interconnect may observe the completion of a read transaction and generate an event packet containing the corresponding Event ID, along with metadata fields that specify the transaction size and completion latency. Similarly, for cache-related events, the Source ID can denote the originating master (e.g., CPU core or DMA engine) that initiated the request. In processor-based EVUs, the Source ID can also encode privilege or process context, for example, RISC-V privilege levels (e.g., user, supervisor, machine) or \new{Arm} Address Space Identifiers (ASIDs)~\cite{arm_armv8a}. 

The physical layer specifies the hardware-level signaling and connectivity between an EVU and its corresponding APMU port. Each EVU is connected to the APMU through a dedicated physical interface, which can be implemented in different forms depending on use cases and area constraints. A straightforward design employs a parallel interface, where the EVU transmits a complete event packet to the APMU every clock cycle over a set of parallel wires. While this approach allows rich event information to be transferred, including fields such as Event Info and Source ID, it may incur significant wiring overhead, especially when the packet width is large. When no event occurs, the EVU transmits a NULL event packet (or an idle code) to maintain timing alignment and ensure deterministic packet framing between EVU and APMU.
Alternatively, the physical layer may employ a \new{multi}-hot encoding,
where each bit in the interface corresponds to a specific Event ID. This approach allows multiple events to be signaled in the same cycle while minimizing encoding complexity. In this case, 
the Event Info and Source ID fields
are either shared across all events or omitted when unnecessary. Designers may also define custom physical interfaces that conform to the logical-layer protocol while optimizing for area, timing, or bandwidth constraints. 

Depending on the required level of flexibility, EVUs may be implemented as either fixed-function or configurable modules. In the former case, the set of architectural events to be monitored is determined at design time, which simplifies implementation and reduces area overhead. Alternatively, the EVU may support configurability at run time, allowing software to dynamically map architectural events to available Event IDs through configuration registers. This enhances observability and adaptability at the cost of potential increases in area and control complexity. In this work, we implement two representative EVU types: 
\new{(i)} a fixed-function AXI4-SPU that uses a parallel interface to capture transaction-level events and latency metadata \new{(Section~\ref{subsec:SPUimpl})}, and \new{(ii)} a configurable CVA6-EVU that exposes a  \new{multi}-hot encoded event interface for core-level micro-architectural events and PC tracking \new{(Section~\ref{subsec:cva6evu})}.

Since EVUs typically operate at the same frequency as the IPs they monitor, clock domain boundaries may exist between the EVU and the APMU. The interface supports clock-domain crossing (CDC) through asynchronous FIFOs, which buffer valid event packets and drop NULL packets. Note that event loss cannot be prevented if the long-term event production rate exceeds the APMU frequency.

\subsection{Advanced Performance Monitoring Unit (APMU)} \label{subsec:apmu}
\begin{figure} [ht]
        \centering
        \includegraphics[width=0.52\textwidth]{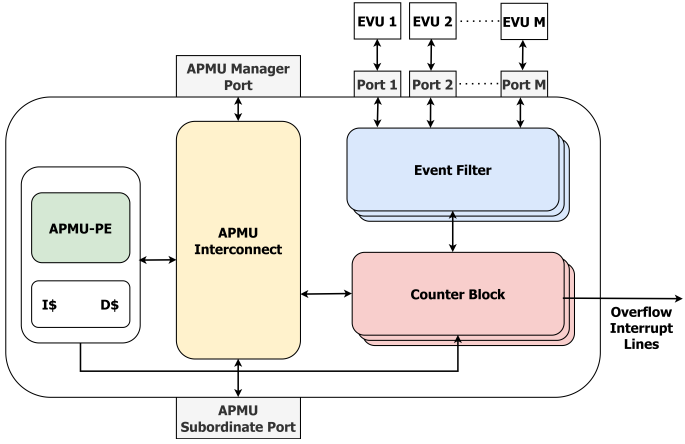}
        \caption{APMU Block Diagram with $M$ EVUs.}
        \label{fig:apmu_emon}
\end{figure}

The APMU is the central component of the infrastructure, collecting event packets from distributed EVUs and performing on-line processing, filtering, and policy-driven actions. It comprises programmable counters that update their state based on incoming events and a lightweight processing element (APMU-PE) with local memory \new{(Instruction and Data ScratchPad Memories (ISPM and DSPM), labeled I\$ and D\$ respectively, as shown in 
Figure~\ref{fig:apmu_emon})}, capable of executing control programs on the collected data, allowing both statistical monitoring and real-time policy enforcement without relying on application cores.

\noindent\textbf{Functional decomposition.}
The APMU exposes two primary capabilities: (1) event aggregation \& selection, (2) programmable counting \& analysis. Event aggregation collects incoming packets from multiple EVU ports and applies programmable selection criteria (Event ID, Source ID, context identifiers) so that downstream logic only processes relevant data. Programmable counters support both simple counting and richer functional modes (conditional increments, adds, min/max, and range tests) on slices of the Event Info field.

\noindent\textbf{Counters \& Programmable Control.}
Each counter can (a)~filter events by Event/Source/Port IDs, (b)~select a bitfield within Event Info for arithmetic or logical evaluation, and (c)~operate in simple-count or functional modes (accumulate, conditional increment/add, keep-min/keep-max). In simple-count mode, the counter increments each time a selected event occurs. In functional modes, the counter can manipulate the Event Info field before updating. For instance, in accumulate mode it adds the Event Info value to the counter, while keep-max/keep-min modes retain the larger or smaller value, respectively. Conditional modes allow selective updates based on comparisons (e.g., equal, less than, in-range), enabling fine-grained event evaluation. Each counter also maintains \textit{overflow} (indicating wrap-around or saturation) and \textit{pending} status flags; the latter is asserted whenever the counter is updated. These primitives, combined with the APMU-PE, allow software to implement flexible monitoring policies, such as per-task accumulation, latency histograms, bandwidth regulation.

\noindent\textbf{Event Processing.}
The APMU operates on an event-driven execution model, where event packets generated by distributed EVUs are streamed into the central monitoring unit. Each incoming event is compared against the filter configuration of one or more counters, and when a match is found, the corresponding counters update their state according to the selected functional operation. This design decouples the counter functionality from the application execution on the APMU-PE, enabling event processing to occur transparently in hardware.

\noindent\textbf{Wait Semantics.}
The APMU-PE supports synchronous event reactions through Wait-for-Pending and Wait-for-Overflow states, blocking until selected counters signal activity. When a counter status flag transitions, the APMU-PE resumes execution with a bitmap of the triggering counters, enabling immediate, context-specific handling. This event-loop model reduces software overhead and allows low-latency responses to events, such as bandwidth throttling, fine-grained statistics collection, or runtime monitoring reconfiguration.

\noindent\textbf{APMU System Control Mechanism.} The APMU’s control capabilities stem from its system-level access and authority. It can influence the platform through two primary mechanisms. First, 
the APMU can raise interrupts to the application cores via the platform interrupt controller based on the status flags in each counter. 
Interrupts allow the APMU to influence system behavior, for example, 
triggering throttling actions \new{to enforce real-time resource regulation (see Section~\ref{sec:regulation}).}
Second, the APMU-PE can read from and write to other components in the system through the shared interconnect \new{via its manager port}. This enables the APMU to not only report status or summary data (for example, writing counter snapshots or compact event records to a log region) but also to dynamically configure or interact with parts of the system in response to monitored conditions. Together, these capabilities make the APMU an active control agent.

\noindent\textbf{Application Core-APMU Interaction.} Communication between the APMU and the application cores follows a memory-mapped model over the system interconnect. Software running on the application cores can program event selectors, map Event IDs, and configure counter behaviour by writing to the APMU configuration registers, which are also accessible to the APMU-PE, allowing its control program to dynamically reconfigure counters or event mappings during runtime. 
Beyond configuration, the communication is bidirectional. The application cores can read and write both the APMU counters and the local memory \new{through the APMU's subordinate port} associated with the APMU-PE, enabling them to inspect monitoring data or update control parameters at runtime. 
The program image of the APMU-PE can be updated at runtime by writing a new binary into that memory over the system interconnect. Execution control of the APMU-PE, such as start, halt, or resume, is managed through a small set of control registers exposed in the same memory-mapped interface.

\subsection{APMU Software Stack}\label{sec:sw_stack}

\new{Integrating the APMU into an embedded system requires a software stack that lets developers write monitoring code that runs efficiently on the resource-constrained APMU-PE, and that interoperates with the OS or hypervisor on the application cores. We therefore designed the stack around three goals: (i) make the APMU's custom instructions usable from a standard high-level toolchain, (ii) provide an event-driven runtime that lets monitoring logic react to hardware events without involving the application cores, and (iii) expose a controlled interface that lets an OS or hypervisor install, configure, and remove monitoring components at run time.}

\new{\noindent\textbf{Load-and-Run Programming Model.}
As mentioned in Section~\ref{sec:design}, monitoring code on the APMU-PE is authored as a set of self-contained \emph{components} rather than as application-side code. A component carries its own \texttt{init} and \texttt{exit} hooks and registers two callbacks with the runtime: an \emph{event handler}, invoked when one of the component's monitored counters signals activity, and a \emph{request handler}, invoked when a peer (typically an application core) sends it a service request. Components compiled into the APMU image are \emph{static} and installed at boot; components shipped at run time (e.g., by a tenant that needs a temporary regulator or tracer) are \emph{dynamic} and installed or torn down through the integration paths discussed below. An internal Base Component owns the request queue and routes incoming requests to the appropriate component by Request ID, isolating component logic from the queue mechanics.}

\new{\noindent\textbf{Event-Triggered Runtime Model.} The APMU-PE runtime is a single non-preemptive event loop built directly on the Wait-for-Pending (WFP) custom instruction. At installation, each component registers its \emph{event bitmask}, the set of counters whose activity should invoke its event handler, and the runtime accumulates these into a global bitmask formed by the logical OR of all registered components. The loop issues WFP on this global bitmask, and the APMU-PE blocks in a low-power state until any monitored counter asserts its pending flag. WFP then returns a bitmap of the counters that triggered the wake-up; the runtime ANDs this bitmap against each component's bitmask and dispatches the event handler of every component with a nonzero match, passing the bitmap so the handler can resolve which counters fired.}

\noindent\textbf{Application/OS Integration.}
The proposed infrastructure is designed to support integration with an Operating System (OS) running in privileged mode. Address translation through the Memory Management Unit (MMU) on each application core can be used to prevent user processes from accessing the APMU and the EVU configuration registers, and from tampering with the code and data of the APMU-PE program. However, to facilitate communication between an application and the APMU, the OS can allow a specific process access to one or more APMU counters, or to a portion of the APMU data local memory, by mapping specific physical memory pages into the virtual address space of the process. \new{For example, this might allow the application to directly notify the APMU-PE of a software event of interest by writing to a counter, to read profiling information generated by the APMU-PE, or to issue a service request to a component using that components \emph{request handler}. Since the} APMU counters are spaced out at page size boundaries in physical memory, the OS can map an individual counter to a given process \new{without exposing the rest of the APMU}. 
\new{Realizing this on a general purpose OS such as Linux involves a similar approach to \texttt{perf}: a kernel module would own the APMU and enforce which counters and memory regions each process can access, while a user-space library would expose the required API.}

\new{\noindent\textbf{Hypervisor Integration.}}
Finally, the same design can also support a virtualized environment, in which case the APMU should be placed under the control of the hypervisor. The two-stage translation mechanism in the MMU can now be used to map specific APMU counters and APMU-PE local memory pages to selected Virtual Machines (VM); in this way, the guest OS for the selected VM can directly communicate with the APMU without requiring hypercalls. Similarly, the IOMMU can be used to prevent a malicious VM from using a DMA to access the APMU memory. We envision a boot process
where the hypervisor loads an APMU image in the APMU local memory and then initializes the APMU-PE. \new{For untrusted guests that should not hold direct access to the APMU, the hypervisor instead exposes a hypercall API that lets a VM install or remove dynamic components. }

\new{\noindent\textbf{Security.}}
Since the APMU can read and write to any memory location, it could \new{tamper} with the memory of the firmware/ secure monitor running at a higher privilege level than the OS (e.g., machine mode in RISC-V or EL3 in \new{Arm}). For this reason, our architecture in Figure~\ref{fig:mc_plat} includes an I/O Physical Memory Protection (IOPMP) unit, acting as a hardware checker to prevent unauthorized access to secure memory. Similarly, we can employ I/O Memory Management Units (IOMMU) to protect system memory against malicious DMA traffic; both Arm and RISC-V have corresponding specifications~\cite{arm_io_mmu, riscv_io_mmu,riscv_io_mmu_ppt}.

\new{\noindent\textbf{Comparison with Conventional Software Stacks.}
The above design differs from the Linux \texttt{perf\_event}/\texttt{perf} and PAPI software paths. All event filtering, counter aggregation, and policy logic runs on the APMU-PE and is woken by hardware, rather than running on an application core under timer interrupts or kernel traps. Consequently the application cores see no per-sample overhead, irrespective of the effective sampling rate. The application cores only involvement in the regulation/monitoring loop is in loading the policy into the APMU-PE's instruction scratchpad.} 

%% file: sections/04_implementation.tex
\section{Implementation} \label{sec:impl}

To validate the proposed architecture and demonstrate its generality, we implemented a complete RTL prototype of the centralized APMU infrastructure in SystemVerilog, including the APMU and multiple EVUs on a RISC-V SoC platform featuring the CVA6 application core~\cite{cva6}. The prototype is designed around standard AXI4 interconnects~\cite{pulp_axi}. We implemented two types of EVUs. 

\new{The SPU} is a generic EVU that monitors AXI traffic between any manager–subordinate pair without altering functionality. Leveraging the wide adoption of AXI4 across commercial and open-source systems (e.g., \new{Arm}, RISC-V, and custom SoCs), the SPU provides a natural observation point for system activity, capturing interactions among multiple IPs without modifying their logic. \new{To demonstrate that our architecture can also leverage existing hardware monitoring support, 
we connect the CVA6's native performance monitoring unit to the APMU through a lightweight EVU wrapper, the CVA6-EVU, without 
redesigning the underlying monitoring logic.}

\subsection{AXI4 Snooping Unit}
\label{subsec:SPUimpl}

\begin{figure} [ht]
        \centering
        \includegraphics[width=0.5\textwidth]{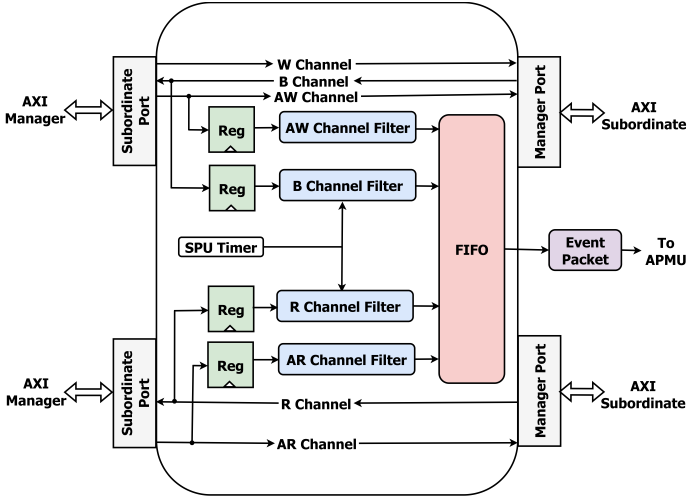}
        \caption{SPU Block Diagram.}
        \label{fig:spu_emon}
\end{figure}

The \new{SPU} monitors AXI4 traffic by routing the original bus through its logic, as shown in Figure~\ref{fig:spu_emon}. It tracks a set of events corresponding to transaction-level activity. In particular, read transactions are tracked through the address read (AR) and read data (R) channels, while write transactions are tracked through the address write (AW), and write response (B) channels (the write data (W) channel is not monitored in our implementation). Request events are identified from address channels (AR/AW), whereas response events are derived from the corresponding data and response channels (R/B). Each event is assigned a unique event ID and can carry additional metadata, such as transaction size, alignment, or response latency. To capture timing-related information it measures the latency of read and write responses by tracking the time between the request issuance and the corresponding response completion. Each event type in the SPU is handled by a dedicated pipeline. The resulting event packets are stored in a FIFO queue, which forms the output interface to the APMU. 

A summary of the monitored events and their associated metadata is provided in Table \ref{table:spu_encoding_emon}. 
Through its parallel interface, the SPU delivers events with latency-sensitive information to the APMU in real time, enabling accurate monitoring. To accurately compute response latency and maintain correct ordering for out-of-order AXI4 transactions with reused IDs, the SPU employs a modified content-addressable memory (CAM) that creates entries for each new request and removes them upon completion.

\setlength{\tabcolsep}{6pt} \begingroup
\renewcommand{\arraystretch}{\tablestretchparam}
\begin{table}[ht]
\centering
\begin{tabular}{lc}
\hline
\textbf{Event} & \textbf{Event Info} \\
\hline
No event & NA \\
\hline
Read request & \multirow{2}{*}{Size of transaction, unaligned transfer} \\
Write request & \\
\hline
Read response & \multirow{2}{*}{Request latency, clock spent in contention} \\
Write response & \\
\hline
\end{tabular}
\caption{AXI4 SPU Event Table}
\label{table:spu_encoding_emon}
\end{table}
\endgroup

\subsection{CVA6 Event Unit (CVA6-EVU)} \label{subsec:cva6evu}

The RISC-V CVA6 core~\cite{cva6} is used as our application core. 
\new{The CVA6 core exposes its native PMU signals as micro-architectural event outputs. Our lightweight EVU wrapper simply interfaces with such signals, selects a subset of events, and packages them into EVU-APMU event packets. We also show that similar logic can be employed to interface with existing pipeline signals to track executing instructions. The CVA6-EVU logic is runtime-configurable, allowing 
software to dynamically select which events and instructions to 
monitor.} 

\noindent\textbf{Event ID Matching and Routing.} This logic continuously monitors a set of incoming performance event signals sourced from the CVA6 hardware performance counters. Each event is uniquely identified by an Event ID. A multi-channel MUX-based selector enables the EVU to track multiple programmable events in parallel. Bitwise matching logic compares incoming signals against these Event IDs, asserting a dedicated status wire for each detected event. This design allows concurrent monitoring of a set of architectural events, including branch mispredicts, pipeline stalls, and exceptions. To extend monitoring to application-level progress, a custom Event ID is reserved for PC markers tracking. 

\noindent\textbf{PC Markers.} This block interfaces with the commit PC and commit acknowledge signals from the CVA6 pipeline. For each committed instruction, the EVU compares the commit PC against a programmable set of reference addresses corresponding to key program locations (e.g., function entry points, basic-block boundaries, or critical code regions). When a match occurs, a dedicated progress-hit signal is asserted. The logic is designed to support multiple commit instructions. Tracking these PC markers enables fine-grained observation of application progress and phase behavior at run time; we demonstrate precise counter attribution in Section~\ref{sec:regulation}.

Detected performance events and PC marker statuses are transmitted via event packets, each consisting of three logical fields: Event ID, Event Info, and Source ID. The EVU employs a \new{multi}-hot encoding scheme at the physical interface, enabling multiple event signals and PC markers to be conveyed in a single packet per cycle. 
The implementation employed in our evaluation supports four event signals and four milestone addresses.
The four performance events are encoded in the Event ID field, while the four PC marker bits occupy the lower portion of the Event Info field. This compact encoding ensures high-throughput, low-latency monitoring without requiring multiple packets.

\subsection{Advanced Performance Monitoring Unit (APMU)}

We employ the Ibex RISC-V core~\cite{ibex} as our APMU-PE,  
configured as a 32-bit RV32IMC processor with a two-stage pipeline (IF and ID/EX), no instruction cache or branch predictor, and a multi-cycle arithmetic unit that takes 3-4 cycles for multiplication.We employ dedicated instruction and data scratchpad memories (ISPM and DSPM) for local code execution and data storage. Instruction fetches from the ISPM complete in one cycle, while LSU accesses to the DSPM take two cycles. To support the monitoring functionality, the APMU-PE extends the base RISC-V ISA with a small set of custom control instructions. The Counter Read and Counter Write instructions provide direct access to APMU counters through a dedicated datapath, completing in two cycles without using the memory interface. 
The Wait-for-Pending and Wait-for-Overflow instructions monitors 
up to 32 counters at a time, corresponding to the processor register width. Note that this restricts 
the number of APMU counters to a maximum of 32. These instructions are implemented using the previously unused opcode \lstinline{7'000_0111}, with distinct function codes assigned for each operation.

Internally, the APMU uses an AXI4-Lite crossbar, adapted from the PULP AXI IP library \cite{pulp_axi}, to connect the APMU-PE, local memories, and counter banks. The crossbar also provides external connectivity through both manager and subordinate AXI4-Lite interfaces. The subordinate port allows system software to configure counters and memory regions, while the manager port enables the PE to initiate 
data transactions.

%% file: sections/05_evaluations.tex
\section{Evaluation}\label{sec:eval}

In this section, we first describe the integration of our developed APMU, AXI4 SPUs, and CVA6-EVUs in a RISC-V-based platform, and report implementation results for an FPGA emulator in terms of resource consumption. Since the emulator runs at low frequency due to FPGA limitations, we also provide frequency results for an ASIC implementation.
We then demonstrate the applicability and benefits of our architecture using two case studies. 
\subsection{Hardware Evaluation Platform} \label{subsec:hw_eval}
\begin{figure}[!t]
    \centering
    \includegraphics[width=\columnwidth]{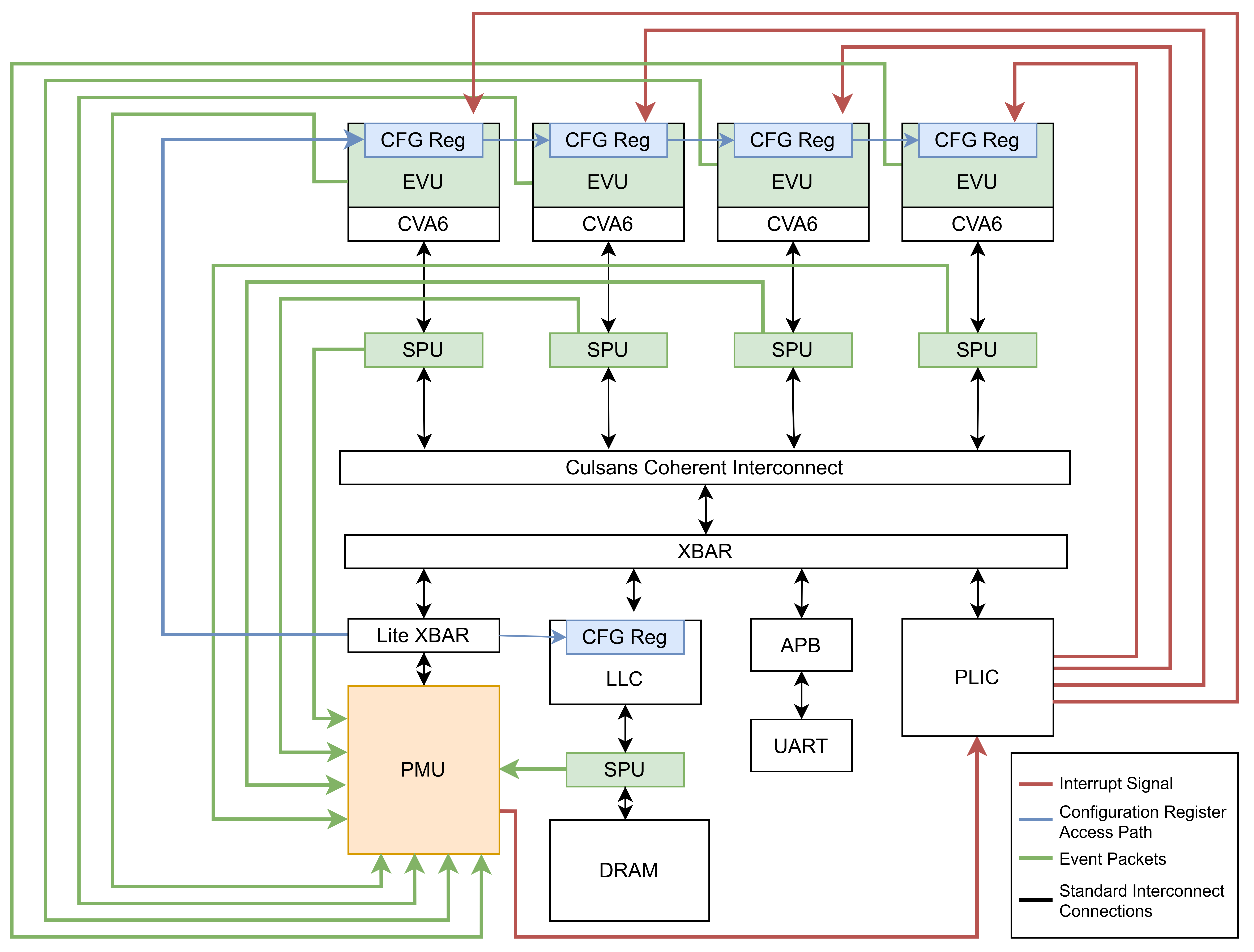}
    \caption{Simplified view of the PULP-based platform.}
    \label{fig:alsaqr}
\end{figure}

For our evaluation, we use a modified variant of the open-source PULP (Parallel Ultra Low Power)~\cite{pulp_org,pulp_website,shaheen_hotchips,shaheen_tcas,shaheen_ojsscs} platform, emulated on the AMD Virtex UltraScale+ FPGA VCU118 board~\cite{vcu118} \new{as shown in Figure~\ref{fig:alsaqr}}. The original PULP platform included a dual-core RISC-V CVA6 processor~\cite{cva6}. 
For our experiments, we extended it to a quad-core configuration, enabling all four cores to execute concurrently. 
Each core has a private 8~KiB 2-way I-Cache and a 16~KiB 8-way D-Cache (16-byte cachelines, write-back policy).
The Culsans unit~\cite{culsans} ensures coherence across the private L1s of the four CVA6 cores. The Culsans manager port is connected to a full AXI4 crossbar, which serves as the primary interconnect fabric of the platform. The crossbar is connected to the PULP platform-level Last-Level Cache (LLC)~\cite{pulp_axi_llc}, which we have extended to support a pseudo-LRU replacement policy and way partitioning, with partitioning registers exposed on the AXI4-Lite crossbar for software-controlled allocation among cores. The LLC, in turn, connects to an off-fabric DRAM. The LLC has a total capacity of 1 MiB, organized as 1024 sets with 16 ways and 64-byte cachelines. It is a write-back, write-allocate cache. The platform employs a debug module based on the SiFive debug specification 0.13 \cite{riscv_dbg}.

Each CVA6 core includes its own internal CVA6-EVUs, whose configuration registers are memory-mapped through the AXI4-Lite crossbar.
Five external AXI4-SPUs have been added: one at the output of each core (core-to-LLC SPU), connected to a subordinate port of the full AXI4 crossbar, and one between the LLC and main memory (LLC-to-Mem SPU). These AXI4-SPUs provide full visibility into memory traffic between the cores and the LLC, and between the LLC and main memory.
The APMU is configured with a 2 KiB ISPM / 16 KiB DSPM and 
is connected to the AXI4-Lite crossbar.

When synthesized to the targeted FPGA device, our APMU implementation can reach a maximum operational frequency of 125 Mhz.
However, due to limitations in the emulation platform, the entire system, including the cores, LLC, EVUs and APMU, is clocked at 20 MHz. Note that in comparison, the off-fabric DRAM runs at 650 MHz. As a result, we acknowledge that most of the inter-core interference for access to main memory arises from how LLC misses are processed in the cache, transferred, and queued in the memory controller, rather than from contention in the DRAM memory device itself. For the same reason, we cannot meaningfully distinguish between contention for access to the same DRAM bank and to different banks~\cite{palloc}.
FPGA resource utilization is presented in 
Table~\ref{table:hw_emon}. \new{The performance counter block, which instantiates the CVA6-EVU, occupies only 0.05\% of LUTs and 0.16\% registers in total, of which the CVA6-EVU itself accounts for 0.02\% of LUTs and 0.04\% registers. The full APMU accounts for 4.49\% of LUTs, of which the counter bank contributes 2.11\% and the APMU-PE 1.08\%.}

\setlength{\tabcolsep}{3pt}
\begingroup
\renewcommand{\arraystretch}{1.15}
\begin{table}[t]
\centering
\small
\begin{tabular}{l r r r r}
\toprule
\textbf{IP Block} 
  & \textbf{LUTs} 
  & \textbf{Regs} 
  & \textbf{BRAM} 
  & \textbf{DSPs} \\
\cmidrule(lr){2-5}
\textit{Platform total} 
  & \textit{773,266} 
  & \textit{342,276} 
  & \textit{6,100} 
  & \textit{115} \\
\midrule
Core Cluster          & 75.90\% & 61.04\% &  6.31\% & 93.91\% \\
\quad One CVA6 core   &  7.15\% &  8.20\% &  1.49\% & 23.48\% \\
\quad\quad Perf.\ Counters &  0.05\% &  0.16\% &     --- &     --- \\
\quad\quad\quad CVA6-EVU    &  0.02\% &  0.04\% &     --- &     --- \\
\addlinespace
APMU                  &  4.49\% &  2.66\% &  0.02\% &  0.87\% \\
\quad Counters \& Regs &  2.11\% &  1.50\% &     --- &     --- \\
\quad APMU-PE         &  1.08\% &  0.70\% &  0.02\% &  0.87\% \\
\addlinespace
One Core-to-LLC SPU   &  0.22\% &  0.47\% &     --- &     --- \\
LLC-to-Mem SPU        &  0.24\% &  0.59\% &     --- &     --- \\
\bottomrule
\end{tabular}
\caption{Resource utilization of the platform and its primary IPs. 
Percentages are relative to the platform total (top row).}
\label{table:hw_emon}
\end{table}
\endgroup

\noindent\textbf{ASIC Implementation Flow.} Our APMU and AXI4-SPU IPs have been integrated in a similar SoC to the one employed for our FPGA emulator, which has been synthesized targeting Global Foundry 22nm node. This dual-core CVA6 implementation employs two AXI4-SPUs connected to the cores, and one to the LLC. The implementation meets a frequency of 800 Mhz for the processor block, which includes the CVA6 cores, Culsans and the two AXI4-SPUs, and 344 Mhz for the rest of the SoC. Based on post-synthesis results, neither the AXI4-SPUs nor the APMU constraint the critical paths in the design. Due to the clock speed difference, CDC FIFOs were inserted on the interface between each AXI4-SPU and the APMU. We validated our integrated IPs based on functional simulations using memory stressor benchmarks (see next Section~\ref{sec:benchmarks}) and empirically found that a FIFO depth of 3 is sufficient to avoid losing events. ASIC utilization is not presented due to flattening of the hierarchy during synthesis. 

\subsection{Software/Benchmarks} \label{sec:benchmarks}

We evaluate our monitoring infrastructure using both synthetic microbenchmarks and practical application benchmarks. All benchmarks are executed on bare metal without an operating system to ensure deterministic execution and eliminate OS-induced scheduling noise. 
We compile code for the CVA6 application cores (RV64IM) and the APMU-PE (RV32IM) 
using the same GCC 15.1.0 toolchain, with appropriate target configurations for each ISA. The APMU-PE custom instructions are exposed to software through inline assembly macros. 

\noindent\textbf{Synthetic Benchmarks.}
We wrote four synthetic benchmarks to stress different levels of the memory hierarchy. 
Each benckmark sequentially reads or writes a contiguous array with a 64-byte stride, ensuring that each access targets a unique cacheline in LLC. The $LLC_r$ stressor performs load operations 
using an array size
larger than the L1 D-Cache size but smaller than the LLC size; thus continuously generating misses in L1 that result in LLC read operations. $LLC_{rw}$ uses the same array size but performs store operations. Due to the write-back L1 policy, it generates both reads and write-backs to LLC. $MM_r$ performs load operations with an array size larger than LLC, thus causing read operations to LLC and to main memory; $MM_{rw}$ performs store operations, generating both reads and writes to LLC and to main memory.

\noindent\textbf{SDVB Benchmarks.}
We also evaluate a set of benchmarks from the San Diego Vision Benchmark (SDVB) suite~\cite{sdvb}, which has been used in prior work on memory regulation and real-time systems, specifically \textit{Disparity}, \textit{MSER}, and \textit{Stitch} as these three workloads exhibit the highest LLC bandwidth among the \new{SDVB} suite~\cite{per-bank}. 
Since we run bare metal without an operating system, benchmark binaries and input data are loaded onto a ramdisk prior to execution. Standard library functions that require system calls (e.g., \texttt{fopen}, \texttt{fread}) are replaced with custom-written wrappers that perform direct memory reads from preloaded datasets. 

\begin{figure}[t]
    \centering
    \includegraphics[width=\columnwidth]{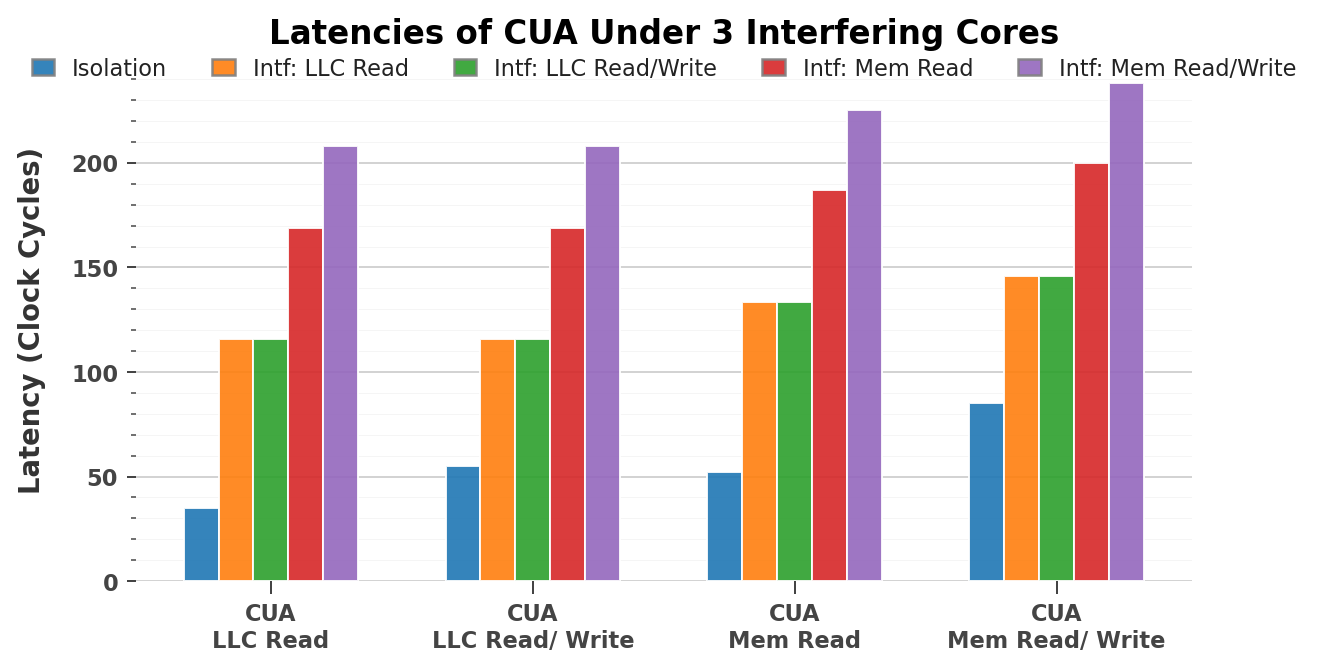}
    \caption{Contention Results: Synthetic Benchmarks. Latency is computed as benchmark execution time divided by the number of load or store instructions.}
    \label{fig:contention_benchmark}
\end{figure}

\begin{figure}[t]
    \centering
    \includegraphics[width=\columnwidth]{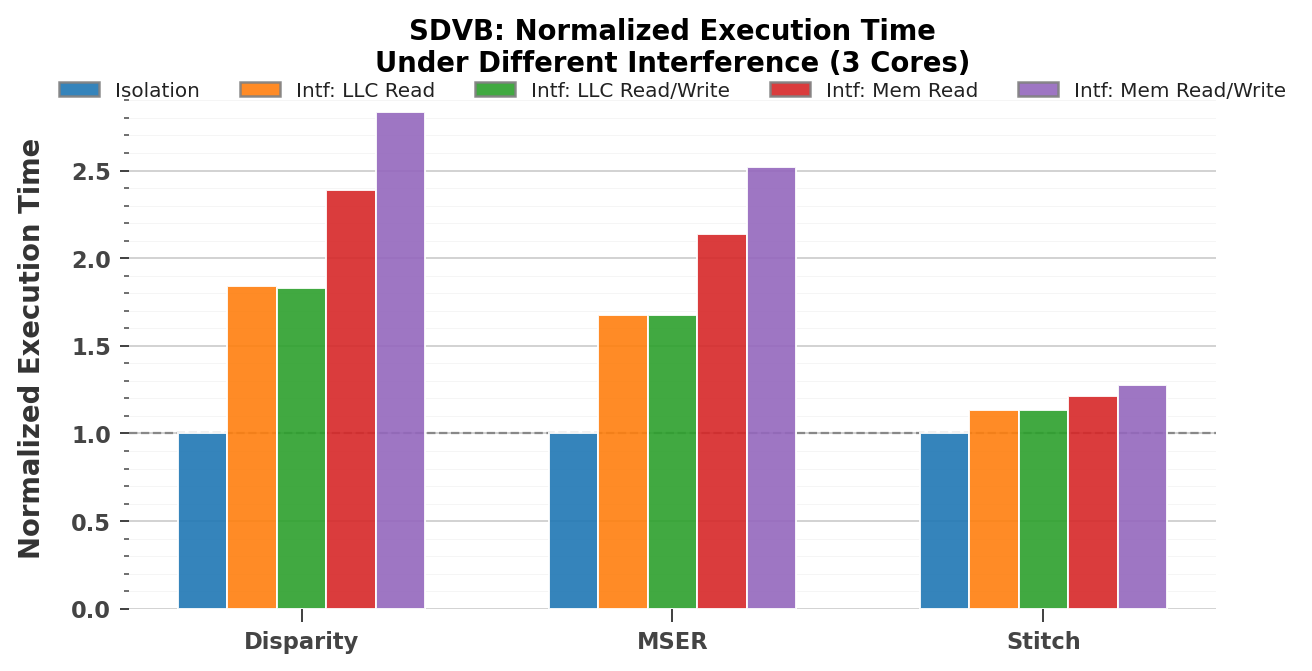}
    \caption{Contention Results: SDVB Benchmarks (sim input). Execution time normalized by the isolation case.}
    \label{fig:sdvb_split1}
\end{figure}

\input{sections/05_regulation.tex}

\input{sections/05_profiling.tex}

\section{Conclusion and Future Work}

We presented a centralized performance monitoring architecture for multicore embedded systems, based on distributed EVUs and a central APMU that decouples event generation from processing with minimal interference on application cores. We implemented and validated the architecture on a quad-core CVA6 RISC-V platform through two case studies: a memory regulation mechanism demonstrating fast control loops with improved execution-time stability, and a fine-grained profiling mechanism enabling precise function-level counter attribution without binary instrumentation. The RTL code of all implemented hardware blocks, the assembled PULP platform and the bare-metal software stack are available open source~\cite{artifact_hesoc, artifact_sw}. \new{Future work includes integrating the software stack with Linux and extending it }
to support virtualized environments, with well-defined interfaces for OS and hypervisor interaction, as well as evaluating area and power overheads in a full ASIC implementation.

%% file: sections/05_regulation.tex
\subsection{Case Study: Memory Regulation} \label{sec:regulation}

In this case study, we evaluate the effectiveness of our monitoring framework in implementing a state-of-the-art software-based memory regulation mechanism. Our implementation uses a polling-based technique inspired by MemPol~\cite{mempol_journal}, where the value of relevant HPCs is read periodically every polling period $P$. 
\new{Note that as discussed in Section~\ref{subsec:HPC}, our RISC-V evaluation platform's HPCs do not support overflow interrupts, making a direct comparison against MemGuard's interrupt-driven mechanism~\cite{memguard} infeasible.}
The implementation in~\cite{mempol_journal} uses the L2 refill and L2 write-back counters, which provide the cumulative number of main memory reads $PMU^r_i(t)$ and main memory writes $PMU^w_i(t)$ performed by core $c_i$ at time $t$. MemPol then computes the \emph{cumulative memory activity} of the core as $A_i(t) = \alpha_r \cdot PMU^r_i(t) + \alpha_w \cdot PMU^w_i(t)$, where $\alpha_r$ and $\alpha_w$ are coefficients that represent the relative impacts of read and write requests on the memory resource. Each core is assigned a budget $A_i^{budget}$, which represents the maximum allowed memory activity per period. At run-time, the MemPol algorithm uses a sliding window approach with a configurable size $w$. If the cumulative memory activity of $c_i$ increases more than $w \cdot A_i^{budget}$ in any window of $w$ polling periods, then the core is halted. Because MemPol only polls the counters and can detect budget overruns 
every $P$ time units,
a core might overshoot its target budget. The second factor that can contribute to overshoot is the delay $D$ between reading the counters and the core being stopped after a budget overrun. Smaller values of $P$ and $D$ decrease the maximum overshoot and thus the amount of time that a core can be halted.

\noindent\textbf{Memory contention in the FPGA emulator.} To determine how to best implement memory regulation, we performed a set of experiments. We run either a SDVB benchmark or one of the four synthetic stress tests on the \emph{core under analysis} (cua). For each benchmark, we run five tests: first, we run the benchmark in isolation with no interfering cores; then, we evaluate the effects of running one of the synthetic stress tests on the other 3 cores. In all tests, we partition the LLC equally among the four cores by assigning 4 private ways (256 KiB) per core; this prevents interfering cores from evicting cache lines of the cua from LLC.

Results in Figure~\ref{fig:contention_benchmark} and Figure~\ref{fig:sdvb_split1} show that the cua can be significantly delayed due to LLC contention: despite LLC partitioning, requests by all cores share internal queues, read/write units, and the hit/miss unit. The $LLC_r$ and $LLC_{rw}$ stressors appear to be causing similar interference, likely due to the parallelism between the read and the write unit in the PULP platform-level LLC. In all cases, interfering main memory traffic causes additional interference over LLC hits, with $MM_{rw}$ stressors causing more interference than $MM_r$: requests that miss in the LLC causing a dirty eviction must be processed sequentially by the LLC evict and refill units~\cite{pulp_axi_llc}.

\noindent\textbf{Algorithm design.} Experimental results reveal that simply capturing the numbers of main memory reads and writes would not be sufficient to precisely capture the effect of interfering core. Instead, we need to also capture the number of LLC reads and LLC writes generated by each core. To this end, we employ 16 counters on the APMU, 4 for each core: 2~counters count the number of either read requests or write requests reported by the LLC-to-Mem SPU, filtering by Source ID to count the requests generated by that core only. We use $PMU^{MM_r}_i(t), PMU^{MM_{w}}_i(t)$ to denote the values of these counters for core $c_i$. The other 2 counters for $c_i$ count the number of either read requests or write requests reported by the core-to-LLC SPU; we use $PMU^{LLC_r}_i(t), PMU^{LLC_{w}}_i(t)$ for their values. We then compute the memory activity as $A_i(t) = \alpha_{LLCr} \cdot PMU^{LLCr}_i(t) + \alpha_{LLCw} \cdot PMU^{LLCw}_i(t) + \alpha_{MMr} \cdot PMU^{MMr}_i(t) + \alpha_{MMw} \cdot PMU^{MMw}_i(t)$. 

Similar to~\cite{memcore}, we employ a token bucket approach:
every polling period, $A_i^{budget}$ tokens are added to the bucket, and a number of tokens equal to the memory activity in that period are removed. The maximum bucket level (number of tokens in the bucket) is $w \cdot A_i^{budget}$. If the bucket level becomes negative, then the core is halted until the level returns non-negative. Since the debug module in our platform cannot be accessed from the crossbar, the APMU regulates the cores by raising interrupts through the Platform-Level Interrupt Controller (PLIC). We use a lightweight ISR that halts the core on a wfi (wait-for-interrupt) instruction and returns to the running program after exiting the wfi.

\noindent\textbf{Computing the coefficients.} In~\cite{mempol_journal}, the coefficients $\alpha_r, \alpha_w$ are computed as follows: first, the authors measured the sustainable memory bandwidth $B_{sustainable}$ of the platform, i.e., the minimum bandwidth (in number of memory requests per unit of time) that can be extracted from all cores running in parallel under worst-case conditions. Each core $c_i$ is assigned a fraction $B_i$ of the memory bandwidth such that $\sum B_i \leq B_{sustainable}$. 
The resulting budget for $c_i$ is $A_i^{budget} = B_i \cdot P$. Since on the platform employed for their evaluation the authors measured the same sustainable bandwidth for read and write requests, they set $\alpha_r = \alpha_w = 1$. 

As shown in Figures~\ref{fig:contention_benchmark}, \ref{fig:sdvb_split1}, reads and writes to LLC and MM have widely different effects in our platform. To compute coefficients $\alpha_{LLCr}, \alpha_{LLCw}, \alpha_{MMr}, \alpha_{MMw}$, we thus use the following approach: we execute on all cores one of the four synthetic benchmarks, and measure the cumulative bandwidth of LLC reads, LLC writes, MM reads and MM writes over all cores. We then
compute the coefficients by equating the derived cumulative memory activity for each of the four cases:
\begin{align}
&\alpha_{LLCr} \cdot B_{LLCr}^{LLCr} = \alpha_{LLCr} \cdot B_{LLCrw}^{LLCr} + \alpha_{LLCw} \cdot B_{LLCrw}^{LLCw} \nonumber\\
&= \alpha_{LLCr} \cdot B_{MMr}^{LLCr} + \alpha_{MMr} \cdot B_{MMr}^{MMr} \nonumber\\
&= \alpha_{LLCr} \cdot B_{MMrw}^{LLCr} + \alpha_{LLCw} \cdot B_{MMrw}^{LLCw} \nonumber\\
&\quad + \alpha_{MMr} \cdot B_{MMrw}^{MMr} + \alpha_{MMw} \cdot B_{MMrw}^{MMw}. \label{eq:coefficients}
\end{align}
\noindent{}In Equation~\ref{eq:coefficients}, the subscript in bandwidth term $B$ represents the synthetic benchmark, while the superscript represents the considered LLC or MM operation; recalling from Section~\ref{sec:benchmarks} that each benchmark generates a different set of operations.  
For similarity with~\cite{mempol_journal}, we set $\alpha_{LCCr} = 1$ and then derive the other 3 parameters based on Equation~\ref{eq:coefficients}. The resulting values of the coefficients are provided in Table~\ref{table:mempol_coef}. Per-core bandwidths $B_i$ are assigned such that $\sum B_i \leq B^{LLCr}_{LLCr}$.

\setlength{\tabcolsep}{4pt}
    \begingroup
    \renewcommand{\arraystretch}{1.1}
    \begin{table}[ht]
    \centering
    \small
    \begin{tabular}{lccc}
    \toprule
    $\alpha_{\text{LLCr}}$ & $\alpha_{\text{LLCw}}$ & $\alpha_{\text{MMr}}$ & $\alpha_{\text{MMw}}$ \\
    \midrule
    1 & 0 & 0.612068966 & 0.439655172 \\
    \bottomrule
    \end{tabular}
    \caption{Calculated regulation coefficients.}
    \label{table:mempol_coef}
    \end{table}
\endgroup

\begingroup

\noindent\textbf{Implementation results.} The worst case execution time for one iteration of our implemented APMU control program is 1,039 cycles, which are 3.02 $\mu s$ at the 344 Mhz frequency of the ASIC implementation. This is significantly shorter than the 5.2 $\mu s$ - 2,600 cycles at the 500 Mhz frequency of the R5 side-core - reported in~\cite{mempol_journal}, despite the \new{Arm} R5 core being more performant than the Ibex. While part of the difference is likely due to switching from a sliding window to token bucket regulation, a main reason is the overhead for reading the counters: in~\cite{mempol_journal} the authors report a mean overhead of $274\,ns$ (137 cycles) for reading each counter, compared to just 2 cycles in our platform. Our poll-control delay is similarly small at only 648 cycles, versus 1,510 cycles (3.02 $\mu s$) in~\cite{mempol_journal}. 

\noindent\textbf{Effectiveness of regulation.}
Figure~\ref{fig:regulation_trace} shows a regulation trace for a synthetic benchmark with $B_{ua} = 25\% \; B^{LLCr}_{LLCr}$
(17.2 tokens refilled per iteration). The trace exhibits three phases: LLC writes, then memory writes, then LLC writes again. When memory write activity increases, the number of individual counter events decreases accordingly; yet the weighted average remains constant throughout. This demonstrates that our regulation correctly accounts for the different costs of LLC versus memory traffic. The bottom panel shows token bucket behavior where we observe the token bucket reaches a lowest value of -55 tokens during the most intensive region of the trace. This max overshoot of 55 tokens is less than the total per-period bandwidth budget across all four cores (68.8 tokens), confirming that even in the most intensive region, the overshoot remains bounded within the system's total sustainable bandwidth for a single regulation period. 

Figure~\ref{fig:sdvb_split2} shows the effects of regulation on SDVB benchmarks, again with 25\% bandwidth allocation. We report the execution time of the benchmark in isolation, as well as when running together with 3 synthetic stressors, normalized by the execution time in isolation with no regulation. Over all benchmarks and scenarios, the maximum variation in execution time is 20.7\% compared to 138.9\% of the unregulated case in Figure~\ref{fig:sdvb_split1}. This shows that our approach effectively stabilizes the execution time of a task under analysis independently from the activity of interfering cores.

\begin{figure}
    \includegraphics[width=\columnwidth]{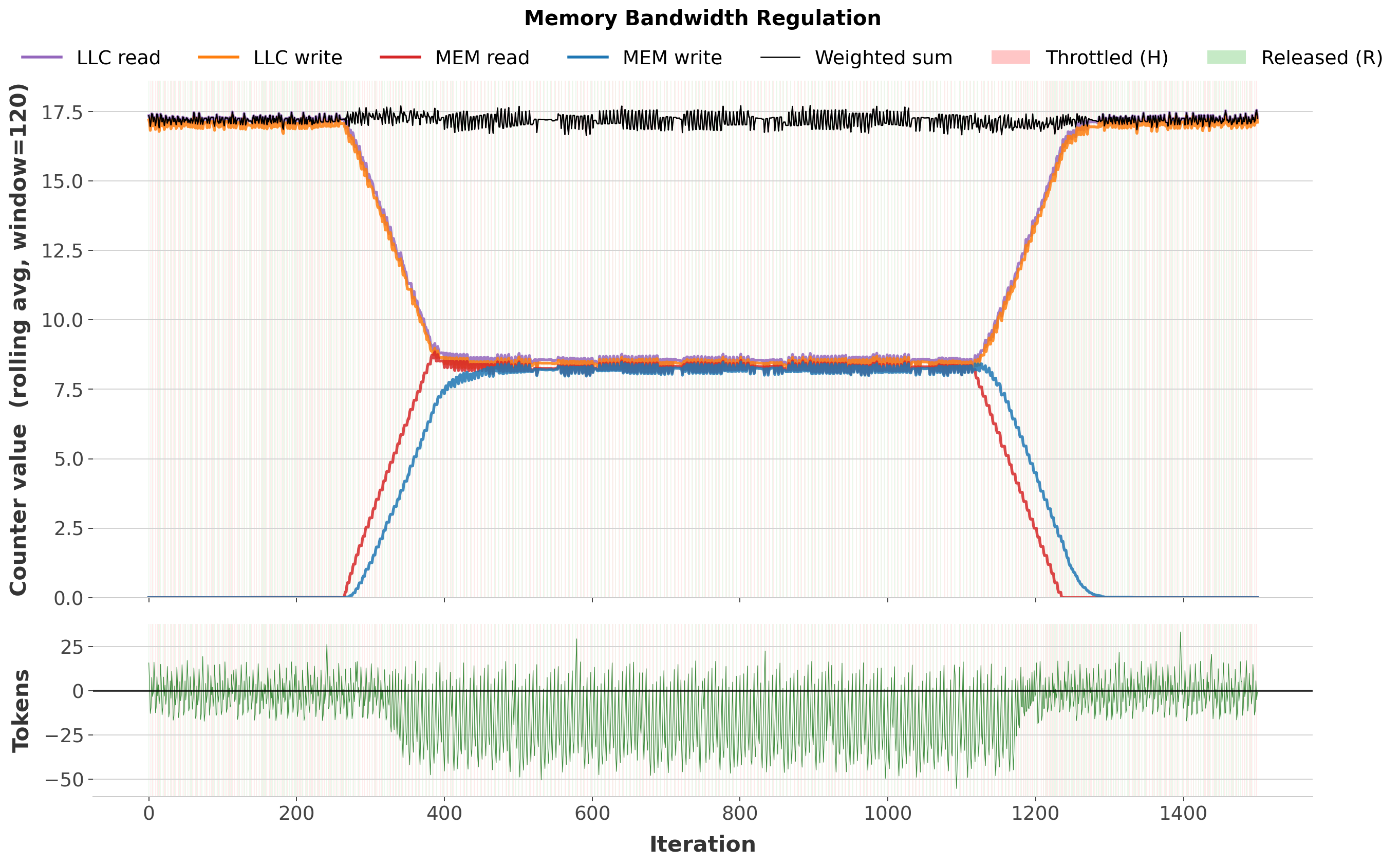}
    \caption{Memory regulation trace under 25\% bandwidth allocation (17.2 tokens/iteration). Top: counter values across three phases (LLC writes $\rightarrow$ memory writes $\rightarrow$ LLC writes). Despite shifting access patterns, the weighted average (black) remains constant. Bottom: token bucket values.}
    \label{fig:regulation_trace}
\end{figure}

\begin{figure}[t]
    \centering
    \includegraphics[width=\columnwidth]{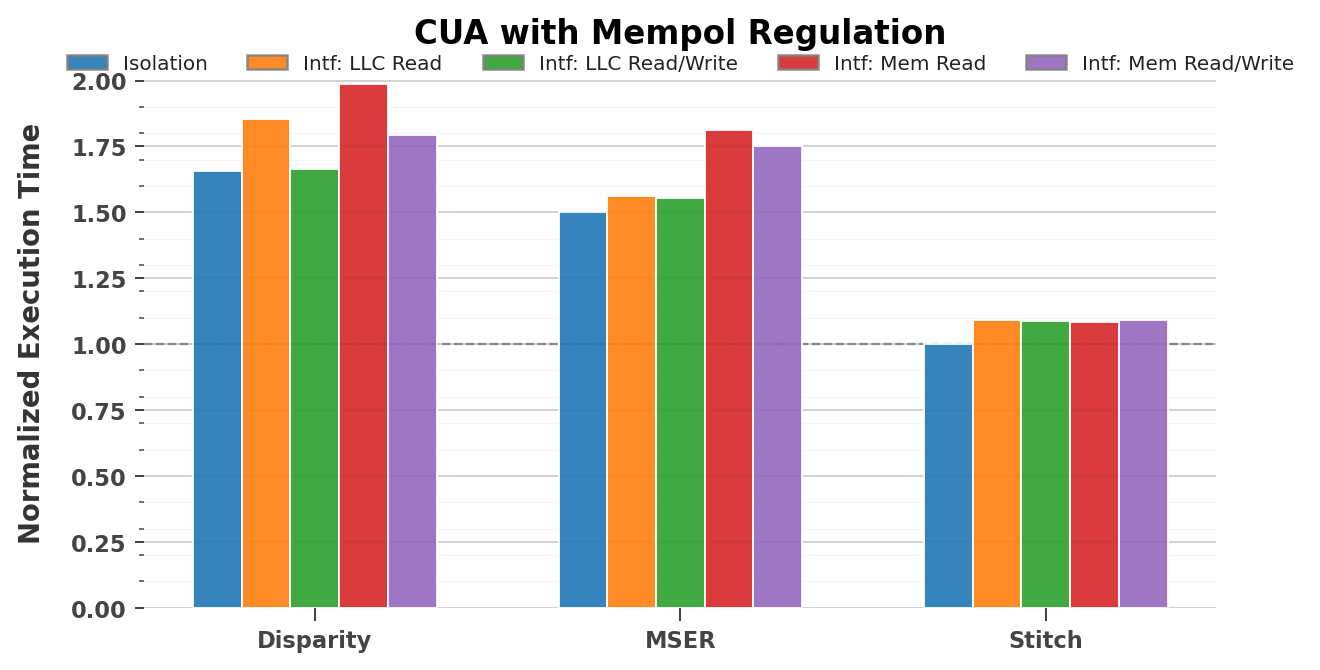}
    \caption{Regulation results: SDVB Benchmarks (sim input). 25\% Bandwidth, $w = 2, P = 2000$ cycles.}
    \label{fig:sdvb_split2}
\end{figure}

%% file: sections/05_profiling.tex
\subsection{Case Study: Profiling and Counter Attribution}
\label{sec:profiling}

This case study demonstrates our framework's ability to attribute hardware performance counters to specific code regions without binary modification, allowing for precise function-level profiling that isolates interference effects at fine granularity with no overhead to the application core.

We profile the SDVB disparity benchmark, which computes a stereo depth map by iterating over 8 disparity levels. Each iteration calls five functions: \texttt{padarray4()} shifts the right image, \texttt{computeSAD()} computes pixel-wise \new{differences}, \texttt{integralImage2D2D()} builds a 2D prefix sum, \texttt{finalSAD()} extracts window sums, and \texttt{findDisparity()} updates the disparity map. We target these functions to understand how microarchitectural behavior varies across them and how memory interference from co-running cores affects each.
\begin{table*}[!t]
\centering
\caption{Totals of function-level hardware counters with and without memory interference (L1D hit ratio is the mean).}
\label{tab:func_counters}
\footnotesize
\setlength{\tabcolsep}{4pt}
\begin{tabular}{l rrrrrrr rrrrrrr}
\toprule
& \multicolumn{7}{c}{\textbf{No Interference}} & \multicolumn{7}{c}{\textbf{3 Cores Interference}} \\
\cmidrule(lr){2-8} \cmidrule(lr){9-15}
\textbf{Function} & Cycles & Loads & Stores & L1D Hit & LLC\_RD & LLC\_WR & RdLat & Cycles & Loads & Stores & L1D Hit & LLC\_RD & LLC\_WR & RdLat \\
\midrule
Padarray4        & 0.80M & 25.7K & 25.2K & 91.4\% & 15.3K &  9.2K & 0.43M & 3.02M & 25.7K & 25.2K & 91.4\% & 15.0K &  9.1K & 2.62M \\
ComputeSAD       & 1.59M & 70.5K & 35.3K & 94.9\% & 25.7K & 10.1K & 0.72M & 5.22M & 70.5K & 35.3K & 94.9\% & 25.7K & 10.1K & 4.34M \\
Integral2d2d     & 1.79M &  105K & 69.8K & 96.6\% & 23.7K & 11.5K & 0.66M & 4.88M &  105K & 69.8K & 96.6\% & 23.7K & 11.6K & 3.73M \\
finalSAD         & 1.23M &  124K & 31.1K & 97.2\% & 15.0K & 10.2K & 0.43M & 3.47M &  124K & 31.1K & 97.2\% & 15.4K & 10.3K & 2.65M \\
FindDisp         & 1.17M & 70.8K & 17.2K & 96.8\% & 16.5K &  6.9K & 0.46M & 3.43M & 70.8K & 17.2K & 96.8\% & 16.3K &  7.1K & 2.72M \\
\bottomrule
\end{tabular}
\end{table*}
To enable function-scoped attribution, we program the CVA6-EVU with entry and exit program counters of each target function, which are easily extracted from the compiled binary's symbol table. The APMU-PE uses Wait-for-Pending to block until function entry, reads all counter values, then blocks again until function exit and reads the counters a second time; the difference yields per-function event counts. The core-to-LLC SPU concurrently tracks LLC requests using the same entry/exit deltas. The event info fields capture per-access latency; the APMU is configured to accumulate latency over each function scope. 
To validate this attribution mechanism, we executed synthetic tests with known expected event counts. Measured counter deltas for loads, L1D misses and LLC reads differed by at most 1 from expected values, confirming that the EVU triggers correctly scope counter increments to the intended function.

\noindent\textbf{Analysis.}
Table~\ref{tab:func_counters} summarizes measurements for all five functions with and without interference from three co-running cores. The key insight is a clear separation between stable and inflated metrics. Access counts (loads, stores, L1D hit ratio, and LLC reads/writes) remain virtually unchanged under interference, confirming that the workload's intrinsic memory behavior is unaffected. In contrast, timing metrics increase dramatically: cycles grow 2.7--3.8$\times$ (e.g., \texttt{padarray4()} from 0.80M to 3.02M), and LLC read latency increases 5.6--6.2$\times$ (e.g., \texttt{computeSAD()} from 0.72M to 4.34M cycles).

This separation pinpoints the interference mechanism: slowdowns stem entirely from increased memory latency due to LLC contention, not from changes in memory demand. Such attribution is possible because our hardware delimits measurements precisely to each function's execution window.

\begin{figure}[t]
    \centering
    \includegraphics[width=\columnwidth]{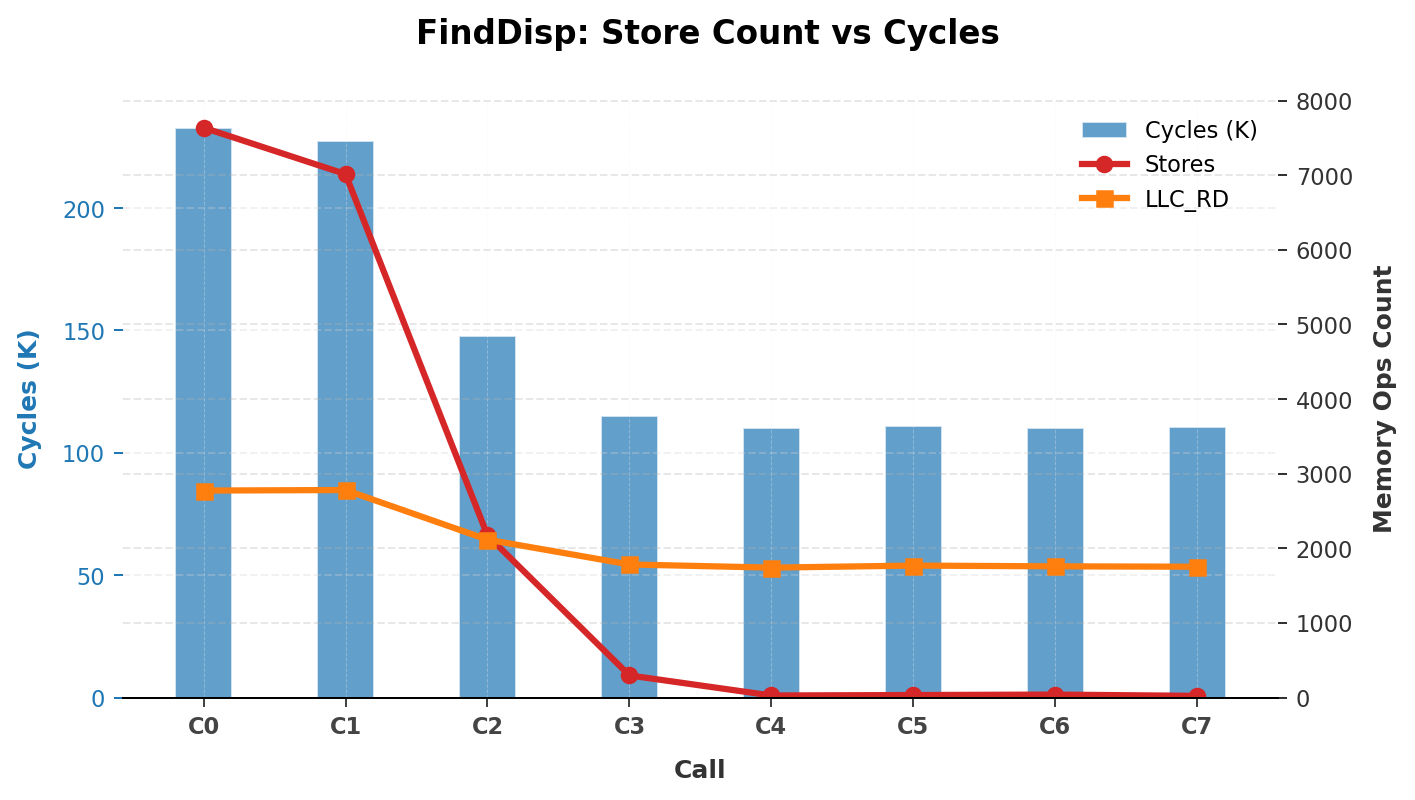}
    \caption{Per-iteration breakdown of \texttt{findDisparity()}. Store count drops sharply after iterations C0--C1, indicating most pixels find their best stereo match early.}
    \label{fig:findDisp_stores_cycles}
\end{figure}

\noindent\textbf{Per-Iteration Insights.}
Beyond aggregate interference analysis, fine-grained profiling reveals algorithmic behavior. Figure~\ref{fig:findDisp_stores_cycles} shows \texttt{findDisparity()} across the 8 disparity iterations. Stores drop sharply after Cycle 1, from $\sim$7.5K to near zero, because the function only writes when a new minimum \new{difference} is found. This indicates most pixels converge to their best stereo match within the first two disparity levels, while later iterations perform comparisons but rarely update the output. Such insights, invisible to coarse-grained profiling, can guide optimizations like early termination.

\new{
\noindent\textbf{Comparison to Software Profiling.} As noted in Section~\ref{section2:related_works}, attributing hardware counters to a specific function on a conventional software stack relies on either interrupt-based hooks, compile-time instrumentation, or periodic polling for statistical profiles, all of which interfere with the application core. As mentioned in Section~\ref{section2:related_works}, polling-based statistical approaches are particularly imprecise: where overflow interrupts are unavailable (as with our platform), periodic polling systematically misattributes events among functions whose execution length is comparable to the polling interval.}

\new{The standard Linux approach for more precise function attribution uses \emph{uprobes}. We run our platform on Linux kernel version 5.10.7 with uprobes backported and measure that placing uprobes at the function entry and exit adds an average latency of $34{,}400$ clock cycles to each profiled function (excluding the actual profiling code) due to the kernel-trap overhead of processing the required system calls. Since the  counters are read inside the handler the events that retire across this gap are a form of interrupt skid, charged to the target function rather than to the trap, so the captured deltas are both inflated and imprecise.}

\new{Another method is GCC's \texttt{-finstrument-functions}, which inserts entry/exit hooks into the compiled binary rather than trapping to the kernel. Since the binary must be modified and rebuilt based on the profiled function(s), we argue that this can be undesirable in a deployed embedded setting.
Furthermore, while the method avoids the kernel-trap overhead, the inserted hooks must still read the counters inside the handler, being susceptible to skid and modifying the timing of the application based on the handler length and the function call frequency.
In contrast, our mechanism described above requires no binary modification. 
The EVU is programmed with entry/exit PCs taken directly from the symbol table and can be retargeted to a different set of functions at runtime, with no changes to the application core and no overhead tied to handler cost or invocation frequency.}

%% file: main.bbl
\begin{thebibliography}{10}

\bibitem{arm_mpam}
Memory system resource partitioning and monitoring ({MPAM}), for {A}-profile
  architecture.
\newblock \url{https://developer.arm.com/documentation/ddi0598/latest/}.

\bibitem{intel_rdt}
{Intel Resource Director Technology (RDT) Framework}.
\newblock
  \url{https://www.intel.com/content/www/us/en/architecture-and-technology/resource-director-technology.html}.

\bibitem{xilinx_perf_ip}
{AMD Xilinx}. {AXI} performance monitor, {Zynq UltraScale+ Device Technical
  Reference Manual (UG1085)}.
\newblock
  \url{https://www.xilinx.com/products/intellectual-property/axi_perf_mon.html}.

\bibitem{profiling_ws_comp}
S.~Kanev et~al.
\newblock Profiling a warehouse-scale computer.
\newblock {\em SIGARCH Comput. Archit. News}, 43(3S):158--169, 2015.

\bibitem{mdelta}
A.-R. Adl-Tabatabai et~al.
\newblock Prefetch injection based on hardware monitoring and object metadata.
\newblock {\em SIGPLAN Not.}, 39(6):267--276, 2004.

\bibitem{maphea}
D.-J.~Oh et~al.
\newblock {MaPHeA}: A framework for lightweight memory hierarchy-aware
  profile-guided heap allocation.
\newblock {\em ACM Trans. Embed. Comput. Syst.}, 22(1), 2022.

\bibitem{memguard}
H.~Yun et~al.
\newblock {MemGuard}: Memory bandwidth reservation system for efficient
  performance isolation in multi-core platforms.
\newblock In {\em RTAS}, pages 55--64, 2013.

\bibitem{memguard_dos}
M.~G. Bechtel and H.~Yun.
\newblock Denial-of-service attacks on shared cache in multicore: Analysis and
  prevention.
\newblock {\em CoRR}, abs/1903.01314, 2019.

\bibitem{memguard_kvm}
N.~Dagieu et~al.
\newblock {Memguard}: Memory bandwidth management in mixed criticality
  virtualized systems.
\newblock 2016.

\bibitem{mempol_journal}
A.~Z{\"u}pke et~al.
\newblock {MemPol}: Policing core memory bandwidth from outside of the cores.
\newblock {\em Real-Time Systems}, 2024.

\bibitem{lat_distribution}
A.~Saeed et~al.
\newblock {Memory Latency Distribution-Driven Regulation for Temporal Isolation
  in MPSoCs}.
\newblock In {\em ECRTS}, pages 4:1--4:23, 2023.

\bibitem{BRU}
F.~Farshchi et~al.
\newblock {BRU}: Bandwidth regulation unit for real-time multicore processors.
\newblock In {\em RTAS}, pages 364--375, 2020.

\bibitem{secure_gpu}
S.~Na et~al.
\newblock Common counters: Compressed encryption counters for secure {GPU}
  memory.
\newblock In {\em HPCA}, pages 1--13, 2021.

\bibitem{gpu_llc_cntr}
J.~Barrera et~al.
\newblock Contention tracking in {GPU} last-level cache.
\newblock In {\em ICCD}, pages 76--79, 2022.

\bibitem{LLC_cntr}
P.~Roy et~al.
\newblock Designing secure performance metrics for last level cache.
\newblock In {\em IPDPSW}, pages 383--392, 2023.

\bibitem{LLC_side_channel_cntr}
P.~P. Bhade and S.~Sinha.
\newblock Detection of cache side channel attacks using thread level monitoring
  of hardware performance counters.
\newblock In {\em MCSoC}, pages 210--217, 2021.

\bibitem{Mem_regulation}
A.~Pradhan et~al.
\newblock Predictable memory bandwidth regulation for {DynamIQ} {Arm} systems.
\newblock In {\em RTCSA}, pages 126--137, 2025.

\bibitem{DRAM_error_cntr}
J.~Chen et~al.
\newblock {CARE}: Coordinated augmentation for elastic resilience on {DRAM}
  errors in data centers.
\newblock In {\em HPCA}, pages 533--544, 2021.

\bibitem{perf_website}
Perf wiki.
\newblock \url{https://perf.wiki.kernel.org/index.php/Main_Page}.

\bibitem{overhead_analysis}
T.~R{\"o}hl et~al.
\newblock Overhead analysis of performance counter measurements.
\newblock In {\em ICPPW}, pages 176--185, 2014.

\bibitem{adjustable_overhead}
D.~Lo et~al.
\newblock Run-time monitoring with adjustable overhead using dataflow-guided
  filtering.
\newblock In {\em HPCA}, pages 662--674, 2015.

\bibitem{perf_stat_man}
{Linux man-pages project}.
\newblock \texttt{perf-stat}(1) {Linux} manual page.
\newblock \url{https://www.man7.org/linux/man-pages/man1/perf-stat.1.html},
  2024.
\newblock Accessed: 2026-06-05.

\bibitem{tip}
B.~Gottschall et~al.
\newblock {TIP}: Time-proportional instruction profiling.
\newblock In {\em MICRO}, pages 15--27, 2021.

\bibitem{precise_event}
J.~Yi et~al.
\newblock On the precision of precise event based sampling.
\newblock In {\em APSys}, pages 98--105, 2020.

\bibitem{sok}
S.~Das et~al.
\newblock {SoK}: The challenges, pitfalls, and perils of using hardware
  performance counters for security.
\newblock {\em IEEE S\&P}, pages 20--38, 2019.

\bibitem{arm_coresight}
{CoreSight} architecture.
\newblock
  \url{https://developer.arm.com/Architectures/CoreSight\%20Architecture}.

\bibitem{intel_pebs}
{Intel Corporation}.
\newblock {Intel\textregistered{} 64 and IA-32 Architectures Software
  Developer's Manual, Volume 3 (3A, 3B, 3C \& 3D): System Programming Guide}.
\newblock
  \url{https://cdrdv2-public.intel.com/843836/325384-sdm-vol-3abcd-dec-24.pdf},
  2024.

\bibitem{arm_spe_overview}
{Arm Ltd.}
\newblock {Arm Architecture Reference Manual Supplement}: Statistical profiling
  extension, for {Armv8-A}.
\newblock \url{https://developer.arm.com/documentation/109429/0100/}, 2017.

\bibitem{axi}
{AMBA AXI} and {ACE} protocol specification issue f.b.
\newblock \url{https://developer.arm.com}, 2017.

\bibitem{riscv_vol_1}
A.~Waterman et~al.
\newblock {\em The {RISC-V} Instruction Set Manual, Volume {I}: User-Level
  {ISA}, Document Version 20191213}.
\newblock 2019.

\bibitem{riscv_pmu}
``{Zicntr}'' and ``{Zihpm}'' counters --- {RISC-V} extension.
\newblock \url{https://wiki.riscv.org/display/HOME/Ratified+Extensions}, 2023.

\bibitem{arm_a72}
{ARM Cortex-A72 MPCore Processor Technical Reference Manual}.
\newblock \url{https://developer.arm.com/documentation/100095/0003/?lang=en}.

\bibitem{intel_pmu}
{Intel 64 and IA-32 Architectures Software Developer Manuals}.
\newblock
  \url{https://www.intel.com/content/www/us/en/developer/articles/technical/intel-sdm.html}.

\bibitem{riscv_cbqri}
{RISC-V} capacity and bandwidth {QoS} register interface ({CBQRI}).
\newblock
  \url{https://riscv.atlassian.net/wiki/spaces/HOME/pages/16154769/RISC-V+Technical+Specifications},
  2024.

\bibitem{riscv_ssqosid}
{RISC-V} quality-of-service ({QoS}) identifiers.
\newblock \url{https://wiki.riscv.org/display/HOME/Ratified+Extensions}, 2024.

\bibitem{linux_perf_events}
L.~Torvalds and {Linux Kernel Contributors}.
\newblock Linux kernel perf events directory.
\newblock \url{https://github.com/torvalds/linux/tree/master/kernel/events},
  2024.

\bibitem{papi}
S.~Browne et~al.
\newblock A portable programming interface for performance evaluation on modern
  processors.
\newblock {\em Int. J. High Perform. Comput. Appl.}, 14(3):189--204, 2000.

\bibitem{Vtune}
J.~Reinders.
\newblock {\em {VTune} Performance Analyzer Essentials}.
\newblock Intel Press, 2005.

\bibitem{perfutility}
{Perf Wiki}.
\newblock Tutorial: {Linux} kernel profiling with perf.
\newblock \url{https://perf.wiki.kernel.org/index.php/Tutorial}, 2023.

\bibitem{cray_uncore}
{Hewlett Packard Enterprise}.
\newblock uncore(5) --- uncore performance counters.
\newblock HPE Cray Programming Environment Documentation, version 24.03, 2023.
\newblock
  \url{https://cpe.ext.hpe.com/docs/24.03/performance-tools/man5/uncore.html},
  accessed 2026-06-17.

\bibitem{nsight_systems}
{NVIDIA Corporation}.
\newblock {NVIDIA Nsight Systems}.
\newblock \url{https://developer.nvidia.com/nsight-systems}, 2024.

\bibitem{hiddenpmu}
Y.~Yang et~al.
\newblock Exploration and exploitation of hidden {PMU} events.
\newblock arXiv:2304.12072, 2023.

\bibitem{tintin}
A.~Li et~al.
\newblock Tintin: a unified hardware performance profiling infrastructure to
  uncover and manage uncertainty.
\newblock In {\em OSDI}, 2025.

\bibitem{debug2}
G.~W.~Dunlap et~al.
\newblock {ReVirt}: enabling intrusion analysis through virtual-machine logging
  and replay.
\newblock {\em SIGOPS Oper. Syst. Rev.}, 36(SI):211--224, 2003.

\bibitem{debug3}
R.~O'Callahan et~al.
\newblock Engineering record and replay for deployability.
\newblock In {\em USENIX ATC}, pages 377--389, 2017.

\bibitem{opt1}
S.~Bhatia et~al.
\newblock Lightweight, high-resolution monitoring for troubleshooting
  production systems.
\newblock In {\em OSDI}, pages 103--116, 2008.

\bibitem{opt3}
T.~A.~Khan et~al.
\newblock {DMon}: Efficient detection and correction of data locality problems
  using selective profiling.
\newblock In {\em OSDI}, pages 163--181, 2021.

\bibitem{power3}
Y.~Zhai et~al.
\newblock {HaPPy}: hyperthread-aware power profiling dynamically.
\newblock In {\em USENIX ATC}, pages 211--218, 2014.

\bibitem{Response_Time_Analysis}
D.~Dasari et~al.
\newblock Response time analysis of {COTS}-based multicores considering the
  contention on the shared memory bus.
\newblock In {\em IEEE TrustCom}, pages 1068--1075, 2011.

\bibitem{multicore_contention}
E.~D\'{\i}az et~al.
\newblock Modelling multicore contention on the {AURIX\texttrademark} {TC27x}.
\newblock In {\em DAC}, pages 1--6, 2018.

\bibitem{gcc_instrument}
{GCC Team}.
\newblock Program instrumentation options.
\newblock Using the GNU Compiler Collection (GCC), 2024.
\newblock
  \url{https://gcc.gnu.org/onlinedocs/gcc/Instrumentation-Options.html},
  accessed 2026-06-17.

\bibitem{uprobes_doc}
Srikar Dronamraju.
\newblock Uprobe-tracer: Uprobe-based event tracing.
\newblock The Linux Kernel documentation, 2024.
\newblock \url{https://docs.kernel.org/trace/uprobetracer.html}, accessed
  2026-06-17.

\bibitem{uprobes_lwn}
Jonathan Corbet.
\newblock Uprobes in 3.5.
\newblock LWN.net, May 2012.
\newblock \url{https://lwn.net/Articles/499190/}, accessed 2026-06-19.

\bibitem{caladan}
J.~Fried et~al.
\newblock Caladan: mitigating interference at microsecond timescales.
\newblock In {\em OSDI}, 2020.

\bibitem{dna}
R.~Gifford et~al.
\newblock {DNA}: Dynamic resource allocation for soft real-time multicore
  systems.
\newblock {\em RTAS}, 2021.

\bibitem{dcat}
C.~Xu et~al.
\newblock {dCat}: dynamic cache management for efficient, performance-sensitive
  infrastructure-as-a-service.
\newblock In {\em EuroSys}, 2018.

\bibitem{muticore_WCET}
J.~Nowotsch et~al.
\newblock Multi-core interference-sensitive {WCET} analysis leveraging runtime
  resource capacity enforcement.
\newblock In {\em ECRTS}, pages 109--118, 2014.

\bibitem{side_channel}
J.~Cho et~al.
\newblock Real-time detection on cache side channel attacks using performance
  counter monitor.
\newblock In {\em ICTC}, pages 175--177, 2019.

\bibitem{survey_hardware_based_malware}
C.~P.~Chenet et~al.
\newblock A survey on hardware-based malware detection approaches.
\newblock {\em IEEE Access}, 12:54115--54128, 2024.

\bibitem{intel_tdt}
{Intel Corporation}.
\newblock Intel {Threat} {Detection} {Technology} ({Intel} {TDT}).
\newblock
  \url{https://www.intel.com/content/www/us/en/architecture-and-technology/threat-detection-technology-brief.html},
  2024.
\newblock Accessed: 2026-06-05.

\bibitem{multi_core_mem_util}
A.~Saeed et~al.
\newblock Memory utilization-based dynamic bandwidth regulation for temporal
  isolation in multi-cores.
\newblock In {\em RTAS}, pages 133--145, 2022.

\bibitem{max_contention}
J.~Cardona et~al.
\newblock Maximum-contention control unit ({MCCU}): Resource access count and
  contention time enforcement.
\newblock In {\em DATE}, pages 710--715, 2019.

\bibitem{space_riscv}
N.-J.~Wessman et~al.
\newblock {De-RISC}: the first {RISC-V} space-grade platform for
  safety-critical systems.
\newblock In {\em IEEE SCC}, pages 17--26, 2021.

\bibitem{tracking_mem_contention}
A.~de~Lecea~et al.
\newblock Improving timing-related guarantees for main memory in multicore
  critical embedded systems.
\newblock In {\em RTSS}, pages 265--278, 2023.

\bibitem{tracking_coherence_contention}
R.~Pujol et~al.
\newblock Tracking coherence-related contention delays in real-time multicore
  systems.
\newblock In {\em ACM/SIGAPP SAC}, pages 461--470, 2023.

\bibitem{per-bank}
C.~Sullivan et~al.
\newblock Per-bank bandwidth regulation of shared last-level cache for
  real-time systems.
\newblock {\em RTSS}, pages 336--348, 2024.

\bibitem{memcore}
I.~Izhbirdeev et~al.
\newblock Coherence-aided memory bandwidth regulation.
\newblock In {\em RTSS}, pages 322--335, 2024.

\bibitem{token_bucket}
R.~L. Cruz.
\newblock A calculus for network delay. {I}. network elements in isolation.
\newblock {\em IEEE Trans. Inf. Theory}, 37(1):114--131, 1991.

\bibitem{arm_armv8a}
{Arm Ltd.}
\newblock {Arm Architecture Reference Manual} for {A}-profile architecture.
\newblock \url{https://developer.arm.com/documentation/ddi0487/latest}, 2023.

\bibitem{arm_io_mmu}
{Arm System MMU} support.
\newblock
  \url{https://developer.arm.com/Architectures/System\%20MMU\%20Support}.

\bibitem{riscv_io_mmu}
{RISC-V} technical specifications.
\newblock
  \url{https://wiki.riscv.org/display/HOME/RISC-V+Technical+Specifications}.

\bibitem{riscv_io_mmu_ppt}
{RISC-V IOMMU} architecture overview.
\newblock
  \url{https://open-src-soc.org/2022-05/media/slides/RISC-V-International-Day-2022-05-05-14h10-Perinne-Peresse.pdf}.

\bibitem{cva6}
F.~Zaruba and L.~Benini.
\newblock The cost of application-class processing: Energy and performance
  analysis of a linux-ready 1.7-{GHz} 64-bit {RISC-V} core in 22-nm {FDSOI}
  technology.
\newblock {\em IEEE Trans. Very Large Scale Integr. (VLSI) Syst.},
  27(11):2629--2640, 2019.

\bibitem{pulp_axi}
A.~Kurth et~al.
\newblock An open-source platform for high-performance non-coherent on-chip
  communication.
\newblock {\em CoRR}, abs/2009.05334, 2020.

\bibitem{ibex}
{Ibex}: An embedded 32-bit {RISC-V} {CPU} core.
\newblock \url{https://ibex-core.readthedocs.io/en/latest/}.

\bibitem{pulp_org}
D.~Rossi et~al.
\newblock Energy efficient parallel computing on the {PULP} platform with
  support for {OpenMP}.
\newblock In {\em IEEEI}, pages 1--5, 2014.

\bibitem{pulp_website}
{PULP Platform}.
\newblock {PULP} platform.
\newblock \url{https://pulp-platform.org}, 2014.

\bibitem{shaheen_hotchips}
L.~Valente et~al.
\newblock Shaheen: An open, secure, and scalable {RV64} {SoC} for autonomous
  nano-{UAVs}.
\newblock In {\em IEEE Hot Chips}, pages 1--12, 2023.

\bibitem{shaheen_tcas}
L.~Valente et~al.
\newblock A heterogeneous {RISC-V} based {SoC} for secure nano-{UAV}
  navigation.
\newblock {\em IEEE Trans. Circuits Syst. I}, pages 1--14, 2024.

\bibitem{shaheen_ojsscs}
M.~Sinigaglia et~al.
\newblock A heterogeneous system-on-chip with an 83 {GFLOp/s}, 1.2 {TFLOp/s/W}
  parallel programmable accelerator for real-time on-device learning in
  autonomous nano-{UAVs}.
\newblock {\em IEEE Open J. Solid-State Circuits Soc.}, 2026.

\bibitem{vcu118}
{AMD Virtex UltraScale+ FPGA VCU118} evaluation kit.
\newblock \url{https://www.xilinx.com/products/boards-and-kits/vcu118.html}.

\bibitem{culsans}
R.~Tedeschi et~al.
\newblock {Culsans}: An efficient snoop-based coherency unit for the {CVA6}
  open source {RISC-V} application processor.
\newblock arXiv:2407.19895, 2024.

\bibitem{pulp_axi_llc}
{PULP Platform Contributors}.
\newblock {AXI LLC: A Parameterizable AXI4-Compliant Last-Level Cache}.
\newblock \url{https://github.com/pulp-platform/axi_llc}, 2025.
\newblock Accessed: 2026-03-26.

\bibitem{riscv_dbg}
T.~Newsome et~al.
\newblock {\em {RISC-V} Debug Specification, Document Version 1.0-STABLE}.
\newblock 2019.

\bibitem{palloc}
H.~Yun et~al.
\newblock Palloc: Dram bank-aware memory allocator for performance isolation on
  multicore platforms.
\newblock In {\em RTAS}, pages 155--166, 2014.

\bibitem{sdvb}
S.~K.~Venkata et~al.
\newblock {SD-VBS}: The san diego vision benchmark suite.
\newblock In {\em IISWC}, pages 55--64, 2009.

\bibitem{artifact_hesoc}
M.~S.~Jafri et~al.
\newblock {he-soc}: {Centralized Performance Monitoring on a PULP-based SoC
  (EMSOFT 2026 artifact)}.
\newblock Zenodo, \url{https://doi.org/10.5281/zenodo.21543104}, 2026.

\bibitem{artifact_sw}
M.~S.~Jafri et~al.
\newblock {APMU Software}: {Bare-metal case-study software (EMSOFT 2026
  artifact)}.
\newblock Zenodo, \url{https://doi.org/10.5281/zenodo.21543106}, 2026.

\end{thebibliography}
